\documentclass[
reprint,
superscriptaddress,
 amsmath,amssymb,
 aps,
]{revtex4-2}

\usepackage{graphicx}% Include figure files
\usepackage{dcolumn}% Align table columns on decimal point
\usepackage{bm}% bold math
\usepackage{chemformula}
\usepackage{amsmath,amsthm,amssymb}
\usepackage{gensymb}
\usepackage{hhline}
\usepackage{enumitem}
\usepackage{enumerate}
\usepackage{soul}
\usepackage{xcolor}
\usepackage{xcolor} % for colored text
\begin{document}

\title{Energy density driven ultrafast electronic excitations in a cuprate
   superconductor}

\author{Alessandra Milloch}
\affiliation{Department of Mathematics and Physics, Università Cattolica del Sacro Cuore, Brescia I-25133, Italy}
\affiliation{ILAMP (Interdisciplinary Laboratories for Advanced
Materials Physics), Università Cattolica del Sacro Cuore, Brescia I-25133, Italy}

\author{Francesco Proietto}
\affiliation{Department of Mathematics and Physics, Università Cattolica del Sacro Cuore, Brescia I-25133, Italy}
\affiliation{ILAMP (Interdisciplinary Laboratories for Advanced
Materials Physics), Università Cattolica del Sacro Cuore, Brescia I-25133, Italy}
\affiliation{Department of Physics and Astronomy, KU Leuven, B-3001 Leuven, Belgium}

\author{Naman Agarwal}
\affiliation{Elettra - Sincrotrone Trieste S.C.p.A., Trieste, Italy}

\author{Laura Foglia}
\affiliation{Elettra - Sincrotrone Trieste S.C.p.A., Trieste, Italy}

\author{Riccardo Mincigrucci} 
\affiliation{Elettra - Sincrotrone Trieste S.C.p.A., Trieste, Italy}

\author{Genda Gu}
\affiliation{Condensed Matter Physics and Materials Science Division, Brookhaven National Laboratory,
Upton, New York 11973, USA}

\author{Claudio Giannetti}
\affiliation{Department of Mathematics and Physics, Università Cattolica del Sacro Cuore, Brescia I-25133, Italy}
\affiliation{ILAMP (Interdisciplinary Laboratories for Advanced
Materials Physics), Università Cattolica del Sacro Cuore, Brescia I-25133, Italy}
\affiliation{CNR-INO (National Institute of Optics), via Branze 45, 25123 Brescia, Italy}

\author{Federico Cilento}
\affiliation{Elettra - Sincrotrone Trieste S.C.p.A., Trieste, Italy}

\author{Filippo Bencivenga}
\affiliation{Elettra - Sincrotrone Trieste S.C.p.A., Trieste, Italy}

\author{Fulvio Parmigiani}
\email[]{fulvio.parmigiani@elettra.eu}
\affiliation{Dipartimento di Fisica, Università degli Studi di Trieste, Trieste, Italy}
\affiliation{Elettra - Sincrotrone Trieste S.C.p.A., Trieste, Italy}
%\thanks{*Corresponding author: fulvio.parmigiani@elettra.eu}

\begin{abstract}

\vspace{1em}
Controlling nonequilibrium dynamics in quantum materials requires ultrafast probes with spectral selectivity. We report femtosecond reflectivity measurements on the cuprate superconductor Bi$_2$Sr$_2$CaCu$_2$O$_{8+\delta}$ using free-electron laser extreme-ultraviolet (23.5--177~eV) and near-infrared (1.5~eV) pump pulses. EUV pulses access deep electronic states, while NIR light excites valence-band transitions. Despite these distinct channels, both schemes produce nearly identical dynamics: above $T_c$, excitations relax through fast (100--300~fs) and slower (1--5~ps) channels; below $T_c$, a delayed component signals quasiparticle recombination and condensate recovery. We find that when electronic excitations are involved, the ultrafast response is governed mainly by absorbed energy rather than by the microscopic nature of the excitation. In contrast, bosonic driving in the THz or mid-infrared produces qualitatively different dynamics. By demonstrating that EUV excitation of a correlated superconductor yields macroscopic dynamics converging with those from optical pumping, this work defines a new experimental paradigm: FEL pulses at core-level energies provide a powerful means to probe and control nonequilibrium electronic states in quantum materials on their intrinsic femtosecond timescales. This establishes FEL-based EUV pumping as a new capability for ultrafast materials science, opening routes toward soft X-ray and attosecond studies of correlated dynamics.

\end{abstract}

\maketitle
\section{Introduction}

Ultrafast pump–probe spectroscopy has become a key technique for exploring nonequilibrium dynamics in high-temperature superconductors \cite{giannetti2016ultrafast}, particularly the cuprates, where intertwined orders such as superconductivity, charge-density waves, and pseudogap states compete and coexist \cite{keimer2015quantum}. In these materials, the photoinduced response is typically strongly dependent on the photon energy of the pump pulse, which determines whether collective or single-particle excitations dominate the transient dynamics.
At the lowest photon energies, in the THz range ($\sim$1–30~meV), ultrafast pulses can resonantly drive collective modes such as the Josephson plasma resonance \cite{matsunaga2014light}, infrared-active phonons \cite{hu2014optically, liu2020pump}, or the amplitude (Higgs) mode of the superconducting order parameter \cite{katsumi2018higgs}. 
Such excitations have enabled the observation of transient superconducting-like states even above the equilibrium transition temperature, including optically induced interlayer coherence in underdoped YBa$_2$Cu$_3$O$_{6+x}$ \cite{Kaiser2012}, 
non-equilibrium superconductivity driven by intense far-infrared pulses \cite{Budden2021}, and wavelength-dependent enhancement of superconducting correlations in stripe-ordered cuprates \cite{Casandruc2015}.
 In the mid-infrared (50–200~meV), resonant excitation of lattice vibrations — such as Cu–O stretching modes — induces strong electron–phonon coupling and nonlinear phononic effects \cite{forst2011nonlinear, mankowsky2014nonlinear}, producing transient structural distortions that can enhance or even induce superconductivity \cite{fausti2011light}. At higher photon energies, near-infrared (near-IR, 0.5–1.5~eV) and visible (1.5–3~eV) pulses excite interband and charge-transfer transitions across the Mott or Hubbard gaps \cite{giannetti2016ultrafast, smallwood2012tracking}, injecting high-energy quasiparticles that disrupt the superconducting condensate. These processes enable time-resolved studies of quasiparticle recombination, pseudogap collapse, and charge-order melting \cite{rameau2014photoinduced, graf2011nodal, perfetti2007ultrafast,boschini2018collapse}. Despite the differences in excitation channels, the subsequent relaxation pathways often converge toward similar dynamics characterized by quasiparticle thermalization and gap recovery.  Optical and mid-infrared driving have provided important insight into nonequilibrium quasiparticle dynamics \cite{giannetti2016ultrafast, zhang2020photoinduced}. However, these studies were restricted to valence-band optical channels and did not probe deep-electronic or core-level excitation.

Extreme-ultraviolet (EUV) and soft X-ray photons ($>$10~eV), which access deeper core-level transitions and provide element-specific insight into band structure and many-body interactions, have traditionally been used only as probe pulses in time-resolved ARPES \cite{hellmann2012time, sobota2021angle, yang2015inequivalence,cilento2018dynamics}. Tabletop high-harmonic generation (HHG) sources can deliver ultrashort EUV pulses \cite{sansone2011high,li2020attosecond}, but their typical fluences (on the order of a few \textmu J/cm$^2$) are too low to significantly perturb low-energy collective modes in bulk superconductors. As a result, no HHG-based EUV pumping in superconductors has yet been demonstrated, and most EUV applications remain probe-only. It was only with the advent of ultrashort, high-fluence free-electron laser (FEL) sources that EUV-to-soft X-ray light could be deployed as an effective pump, enabling direct access to nonequilibrium regimes unreachable by tabletop systems.\\
%\textcolor{blue}{\st{In this work, we do not seek to address the microscopic origin of unconventional superconductivity. Instead, our objective is to determine which physical parameter governs the nonequilibrium relaxation of a cuprate when it is driven by electronic excitation. We show that when the excitation channel is electronic, the ultrafast dynamics are governed solely by the macroscopic density of ther absorbed energy, independent of whether electrons are promoted via interband transitions or deep–core photoionization.}}\\
Here, we report the first use of ultrashort EUV pulses as a pump to drive ultrafast dynamics in the cuprate superconductor \ch{Bi2Sr2CaCu2O_{$8+\delta$}} (Bi2212), probed with a 1.5~eV near-IR beam. We do not seek to address the microscopic origin of unconventional superconductivity; rather, our aim is to identify the physical parameters that govern the nonequilibrium relaxation of a cuprate driven by electronic excitation. Our results show that electronic excitations ranging from near-infrared charge-transfer to extreme-ultraviolet core-level transitions lead to convergent ultrafast dynamics governed primarily by the absorbed energy density. This establishes EUV free-electron laser spectroscopy as a powerful frontier tool for quantum materials research. By demonstrating that macroscopically similar dynamics emerge from distinct microscopic electronic pathways, the work broadens the scope of pump–probe experiments on correlated matter. Extending this approach to soft X-ray and attosecond FEL sources will allow direct access to the intrinsic timescales of electronic correlations, positioning FEL-based ultrafast spectroscopy as a paradigm for nonequilibrium studies of complex quantum systems.\\
%Our experiment represents a first step toward clarifying whether the observed dynamics reflect direct coupling to the condensate or are instead mediated by indirect scattering cascades. This question can be addressed by complementing our data with theoretical modeling, such as non-equilibrium dynamical mean-field theory (NE-DMFT) \cite{Georges1996}\cite{Eckstein2009}\cite{Amaricci2015}\cite{Werner2012} or time-dependent density matrix renormalization group (tDMRG) \cite{Schollwock2011} \cite{Kollath2007} \cite{Barmettler2008}\cite{Daley2004} studies focused on cascade thermalization or multi-step relaxation, where an initial non-thermal distribution of high-energy excitations (e.g., electrons coupled to bosons) relaxes via intermediate scattering channels, before reaching thermal equilibrium.\\
The measurements were performed at the TIMER endstation of the FERMI-FEL facility, where the operational conditions provide pulses with excitation fluence sufficient to perturb the superconducting condensate. 
First, EUV-pump/optical-probe experiments were conducted at room temperature by scanning the photon energy of the EUV pump pulses across several core-level resonances, and by varying the FEL fluence. Then, in order to establish the conditions under which extreme ultraviolet excitation could transiently suppress or quench the superconducting state, the experiment was repeated in the superconducting phase, at ~30 K, using a 77 eV pump. %These measurements provide critical insights into the efficiency and selectivity of EUV excitation in cuprates, and help determine whether the energy delivered by the FEL could perturb or fully suppress the superconducting condensate. By systematically comparing the transient reflectivity traces obtained above and below Tc, and as a function of both pump fluence, we aim to identify the threshold conditions for inducing significant non-equilibrium changes in the superconducting state. 
The fluence dependence of the EUV-induced dynamics differs above and below the superconducting transition temperature $T_c$ and, remarkably, in both regimes closely resembles that observed with conventional near-IR excitation, revealing a universal non-equilibrium response that spans a wide range of excitation energies. This universality opens new possibilities for high-energy control of correlated states and extends the reach of pump–probe spectroscopy deep into the extreme-ultraviolet domain.

\begin{figure*}
\centering
\includegraphics[width=18cm]{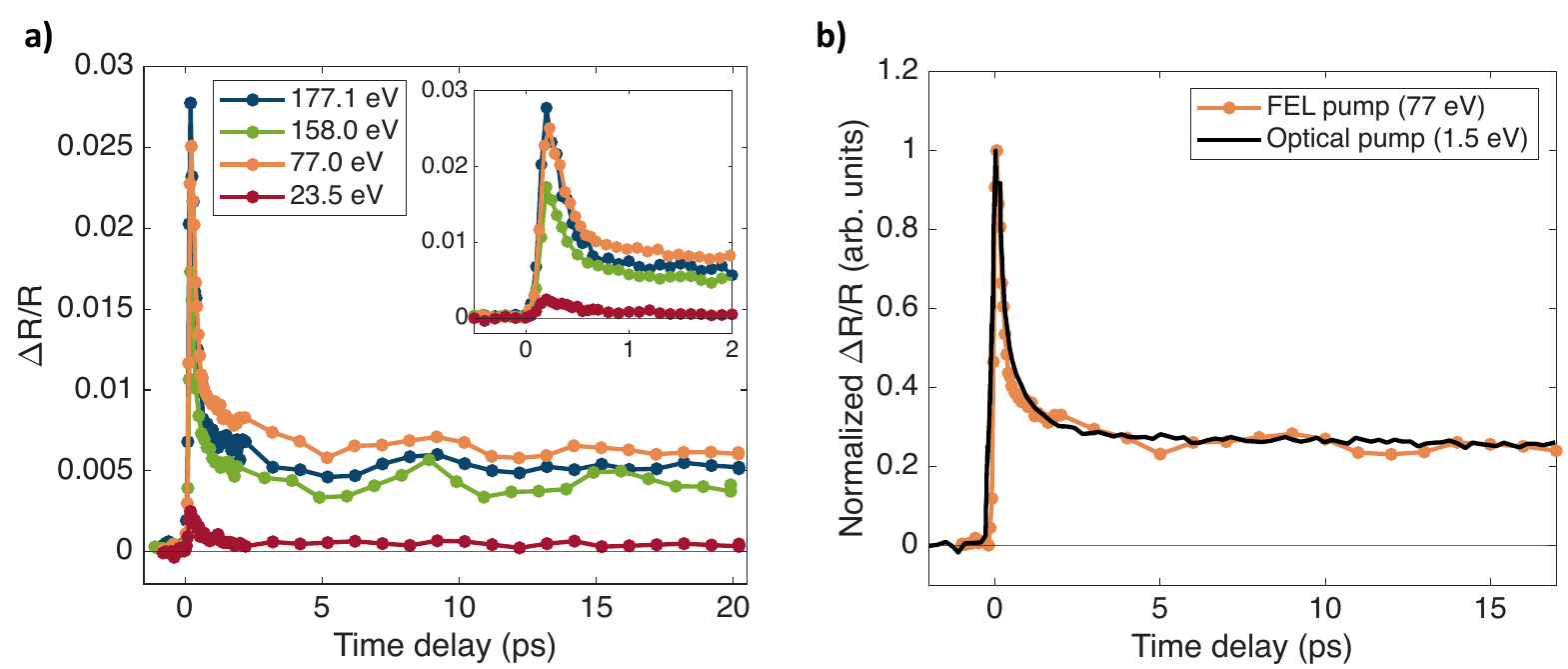}
\caption{Transient reflectivity dynamics at $T = 300$~K. 
(a) EUV-pump, optical-probe measurements for four different pump photon energies $E_{\mathrm{EUV}}$, each taken at an excitation energy density $I_{\mathrm{EUV}} \sim 200$--$300$~J/cm$^3$. The inset shows an expanded view of the first 2~ps of the dynamics. 
(b) Comparison between room-temperature dynamics induced by an EUV pump ($E_{\mathrm{EUV}} = 77$~eV) and by a near-IR pump ($E_{\mathrm{pump}} = 1.5$~eV, adapted from Ref.~\citenum{giannetti2009discontinuity}).}
\label{fig: RT dynamics}
\end{figure*}

\section{Experimental section}

The experiment (schematic in Fig. S1, Supplementary Material \cite{supplemental}) was performed at the TIMER endstation of the FERMI FEL (Elettra Sincrotrone Trieste, Italy). The photon energy of the seeded FEL pulse, $E_{\mathrm{EUV}}$, used as the pump, was tuned between 23.5 and 177.1~eV by operating both the first and second stages of the FERMI FEL2 source \cite{,allaria2015fermi}. The excitation energies were selected to target several core-level resonances; specifically, as summarized in Table \ref{tab:penetration_depths}, $E_{\mathrm{EUV}}$ was tuned across the Cu-3p absorption edge (70.5, 72.1, 75.2, 77.0, and 80 eV), to the O-2s edge (23.5~eV) and the Bi-4f edge (158~eV), as well as off-resonance at 177.1~eV.
The pulse duration was 40–60~fs (FWHM), with higher photon energies corresponding to shorter durations \cite{finetti2017pulse}, and the repetition rate was 25 Hz. The EUV beam was focused onto the sample to a $350 \times 250~\mu$m$^2$ full-width-at-half-maximum (FWHM) spot using a toroidal mirror.
A near-IR probe pulse ($E_{\mathrm{probe}} = 1.58$~eV) was focused to a $250 \times 250~\mu$m$^2$ (FWHM) spot and delayed relative to the pump by a time $\Delta t$. The probe pulse had a 100~fs duration, a 50 Hz repetition rate, and intersected the FEL beam at a $13.8 \degree $ crossing angle. The probe light reflected from the sample was collected by the same focusing lens, separated from the incident probe by a 50:50 beamsplitter, and imaged onto a CCD camera. For each delay $\Delta t$, 1000 single-shot images were recorded. Because the pump repetition rate was half that of the probe, half of the images corresponded to the pump-on condition and half to pump-off. For each image, the CCD counts were first integrated over a region of interest corresponding to the probe spot on the camera. The transient reflectivity change, $\Delta R/R$, was then calculated as the normalized difference between the averaged pump-on and pump-off signals (${\Delta R}/{R} = {(\langle \text{pump-on} \rangle - \langle \text{pump-off} \rangle)}/{\langle \text{pump-off} \rangle}$).

Measurements were carried out on high-quality underdoped \ch{Bi2Sr2CaCu2O_{$8+\delta$}} (Bi2212) single crystals ($\delta = 0.1$), with a superconducting transition temperature $T_c = 70$~K, as determined by SQUID magnetometry. The sample temperature was controlled with an open-loop liquid-helium cryostat.

\begin{table}[htbp]
\centering
\setlength{\tabcolsep}{8pt} % increase column separation
\caption{Pump photon energies used in the experiment, and corresponding core resonances, pump absorption lengths (in nm), and excited-to-probed volume ratios ($V_{\mathrm{exc}}/V_{\mathrm{probe}}$).}
\label{tab:penetration_depths}
\begin{tabular}{cccc}
\hline
\rule{0pt}{18pt} \\[-18pt]
$E_{\mathrm{EUV}}$ (eV) & Resonance & $L_{\mathrm{abs}}^{\mathrm{EUV}}$(nm) & $V_{\mathrm{exc}}/V_{\mathrm{probe}}$ \\
\hline
23.5   & O-2s           & 10.7   & 0.065 \\
70.5   & Cu-3p          & 22.0   & 0.132 \\
72.1   & Cu-3p          & 22.4   & 0.135 \\
75.2   & Cu-3p          & 23.4   & 0.141 \\
77.0   & Cu-3p          & 24.0   & 0.145 \\
80.0   & Cu-3p          & 25.4   & 0.153 \\
158.0  & Bi-4f          & 79.9   & 0.481 \\
177.1  & Off-resonance  & 85.3   & 0.514 \\
\hline
\end{tabular}
\end{table}

In pump–probe experiments combining EUV excitation with near-IR optical probing, the impact of the EUV photon absorption length on the measured response needs to be taken into account. Two key aspects arise from this. 

(i) The volume over which the EUV pump pulse energy is absorbed, and hence the magnitude of the material excitation and its spatial profile, strongly depend on the photon energy, due to the energy-dependent EUV attenuation length ($L_{\mathrm{abs}}^{\mathrm{EUV}}$). To account for this, we estimate the average excitation energy density absorbed over one attenuation length as 
\begin{equation}
    I_{\mathrm{exc}} = (1-e^{-1})F/L_{\mathrm{abs}}^{\mathrm{EUV}},
\end{equation}
where $F$ is the incident fluence. In our experiment, the excitation fluence, $F$, was adjusted between 0.2 and 5~mJ/cm$^2$ using a gas attenuator and thin-film filters (Si, Zr, or Mg); the corresponding energy density $I_{\mathrm{exc}}$ ranged from 50 to 1300~J/cm$^3$. We note that these values were estimated using the nominal transmission of the beamline and should therefore be regarded as an upper limit.
The average lattice temperature increase in the volume within one $L_{\mathrm{abs}}^{\mathrm{EUV}}$ from the surface can then be estimated as $\Delta T_{\mathrm{lat}} =  I_{\mathrm{exc}} /C_{\mathrm{lat}}$, where $C_{\mathrm{lat}}$ is the lattice specific heat; at room temperature $C_{\mathrm{lat}}  \simeq 2.27$~Jcm$^{-3}$K$^{-1}$ \cite{junod1994specific, dal2012disentangling}, resulting in local temperature increase up to 500~K; at low temperature ($C_{\mathrm{lat}}  \simeq 0.22$~Jcm$^{-3}$K$^{-1}$ \cite{junod1994specific}) $\Delta T_{\mathrm{lat}}$ varies between 200 and 5000~K in the investigated fluence range, suggesting a significant role played by heating in the measured transient reflectivity response.
Because $L_{\mathrm{abs}}^{\mathrm{EUV}}$ depends strongly on $E_{\mathrm{EUV}}$, not all values of $I_{\mathrm{exc}}$ were accessible for every photon energy: for the largest $E_{\mathrm{EUV}}$, having longer $L_{\mathrm{abs}}^{\mathrm{EUV}}$, the pulse energy is absorbed over a larger volume, resulting in a lower average excitation magnitude.

(ii) The volume probed by the near-IR probe pulse is much larger than the excited volume, because the near-IR absorption length ($L_{\mathrm{abs}}^{\mathrm{probe}}$) is up to an order of magnitude greater than that of the EUV pump. Due to the mismatch in optical absorption lengths between pump and probe, the recorded reflectivity change represents a spatial average over both excited and unperturbed regions. In our experiment, the 1.5 eV probe absorption length is $L_{\mathrm{abs}} ^{\mathrm{probe}}\approx 166 $~nm, estimated based on equilibrium optical constants of the material \cite{giannetti2011revealing}. In contrast, EUV photons are absorbed within just the first few tens of nanometers of the material. Based on CXRO database values \cite{CXRO}, the pump penetration length in the $E_{\mathrm{EUV}}$ range used here varies between 11 and 85~nm (see Table \ref{tab:penetration_depths}). 
%These values, which are estimated based on atomic scattering factors, are in agreement with tabulated X-ray optical constants for layered cuprates and oxides, that report an absorption coefficient $\alpha_{\mathrm{EUV}} \sim 1.5 \times 10^6$~cm$^{-1}$, giving an absorption length $\sim$ 7~nm. 
Such values, estimated from atomic scattering factors, align with tabulated X-ray optical data for layered cuprates and oxides, reporting an absorption coefficient $\alpha_{\mathrm{EUV}} \sim 1.5 \times 10^{6}\ \mathrm{cm}^{-1}$ (absorption length $\sim$7~nm) \cite{Henke1993}.

Assuming normal incidence and a uniform lateral beam profile, the effective excited volume fraction within the probed region is estimated as
\begin{equation}
    v = \frac{V_{\mathrm{exc}}}{V_{\mathrm{probe}}} \approx \frac {L_{\mathrm{abs}} ^{\mathrm{EUV}} }{L_{\mathrm{abs}} ^{\mathrm{probe}}}.
\end{equation}
As summarized in Table \ref{tab:penetration_depths}, at 23.5~eV pump energy, only $~6.5\%$ of the probed volume is excited, whereas the $V_{\mathrm{exc}}/V_{\mathrm{probe}}$ ratio increases to $\sim15\%$ around 72-80~eV, and to $\sim50\%$ at 158-177~eV. This implies that only a small fraction of the optically probed material, $v$, is directly excited by the EUV pulse, and that the reflectivity signal from the excited volume is diluted by a factor on the order of $\sim1/v$ when averaged over the full probed depth. This penetration-depth mismatch must be accounted for when interpreting the results, which represent spatial averages over the full probed depth. In order to consider the role of the excitation profile decaying exponentially over $L_{\mathrm{abs}}$ inside the material, we model the excitated sample as a layered medium with a depth-dependent refractive index and compute the corresponding reflectivity using the transfer matrix method. As detailed in Supplementary Material  Sec. S2 \cite{supplemental}, this allows us to extract a normalization factor, $r$, for the transient reflectivity amplitude, which depends on the pump absorption length. This factor should be regarded as an approximate, order-of-magnitude correction that facilitates comparison across different pump photon energies, rather than as a precise quantitative measure.%this allows us to extract a normalization factor, $r$, for the transient reflectivity amplitude depending on the pump penetration depth, which should be regarded as an order-of-magnitude normalization to facilitate comparison across different photon energies, rather than a precise quantitative description.}

\section{Results}

\subsection{Room temperature dynamics}
Room-temperature measurements were first performed to assess the influence of the pump photon energy, to characterize the temporal dynamics associated with EUV excitation, and to benchmark the results against established optical-pump/optical-probe measurements.

Figure~\ref{fig: RT dynamics}(a) shows four representative traces measured at room temperature for four different pump photon energies $E_{\mathrm{EUV}}$, each taken at an excitation energy density $I_{\mathrm{exc}} \sim 200$--$300$~J/cm$^3$ (the complete dataset for all employed $E_{\mathrm{EUV}}$ values is provided in Fig. S3 Supplementary Material \cite{supplemental}). The overall temporal profile remains essentially unchanged when tuning the photon energy across different resonances, aside from variations in the transient reflectivity amplitude. Increasing the excitation intensity likewise leaves the shape of the dynamics largely unaffected (see Fig. S3 Supplementary Material \cite{supplemental} for full fluence-dependent measurements).

To quantify these observations and extract recovery timescales, the room-temperature $\Delta R/R$ traces are fitted with a two-exponential decay function,  
\begin{equation}
I_{\mathrm{RT}} (\Delta t) = a_1 e^{-\Delta t / \tau_1} + a_2 e^{-\Delta t / \tau_2} + c,
\end{equation}
multiplied by a Heaviside step function and convoluted with a Gaussian of 120~fs FWHM to account for the pump and probe pulse durations. The dynamics are well reproduced by two components with time constants $\tau_1$ and $\tau_2$, which describe the thermalization of excited electrons with different degrees of freedom, including lattice, strongly coupled phonons, spin and bosonic excitations of electronic origin \cite{,dal2012disentangling}.

\begin{figure}
\centering
\includegraphics[width=8.5cm]{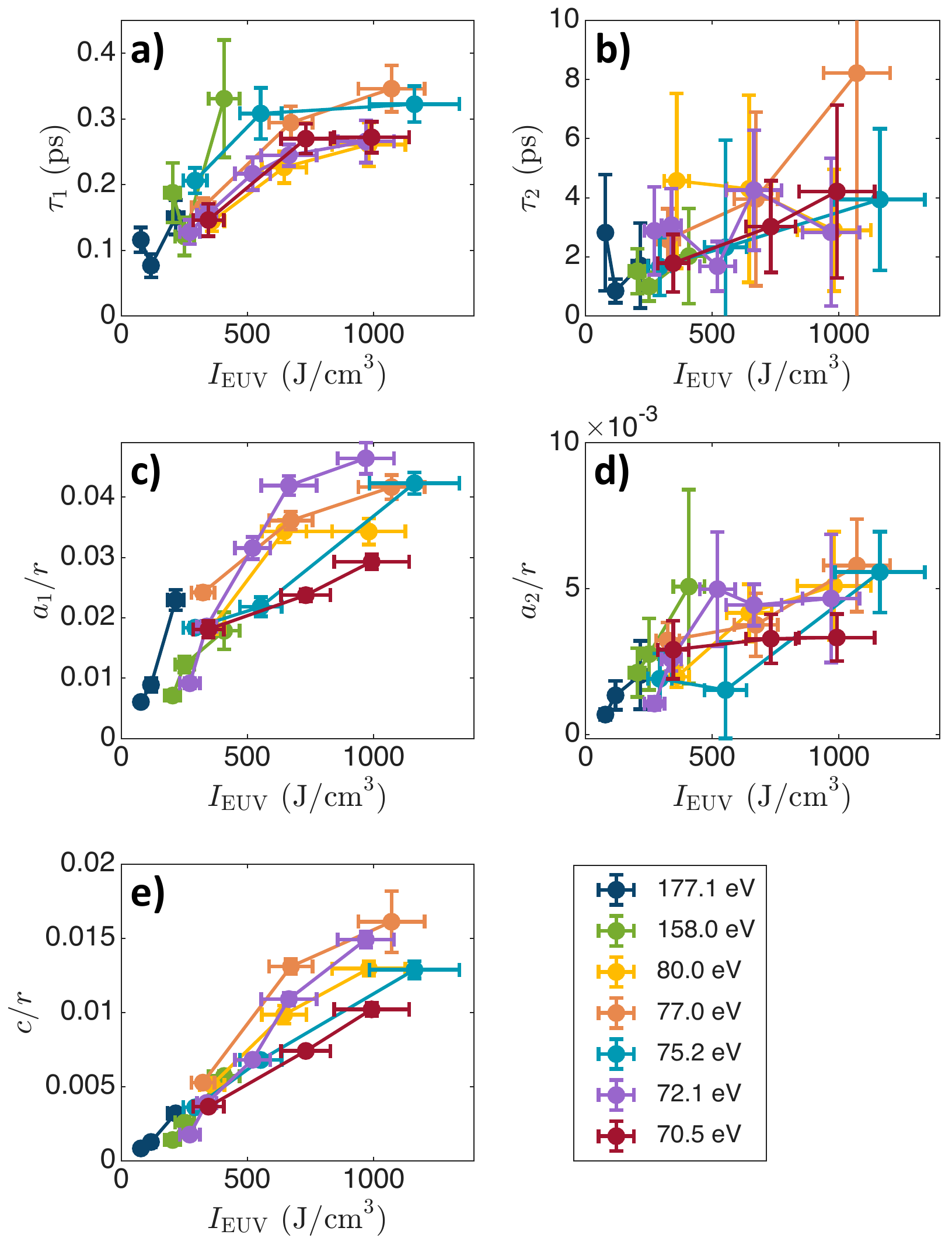}
\caption{Fit parameters extracted from room-temperature transient reflectivity measurements as a function of excitation energy density $I_{\mathrm{EUV}}$ for all pump photon energies employed. 
(a,b) Fast and slow decay times, $\tau_1$ and $\tau_2$, respectively. 
(c,d,e) Amplitude parameters $a_1$, $a_2$, and $c$, each normalized by the factor $r$ extracted from the transfer matrix model (see Supplementary Materials Sec. S2 \cite{supplemental}) to account for penetration-depth mismatch between pump and probe.}
\label{fig: RT fit param}
\end{figure}

A summary of fit results is shown in Fig.~\ref{fig: RT fit param}, where the extracted parameters are plotted as a function of $I_{\mathrm{exc}}$. Data corresponding to $E_{\mathrm{EUV}} = 23.5$~eV (O-2s shallow core level) are not included in Fig.~\ref{fig: RT fit param} because the estimated excitation fluence for this photon energy, resonant with oxygen states, is strongly affected by uncertainties in the beamline transmission. In particular, oxidation of filters and mirrors over time likely reduces the actual transmission, leading to an overestimation of $I_{\mathrm{exc}}$. Fit parameters relative to 23.5~eV pump can be found in Supplementary Material (see Fig. S4) \cite{supplemental}. For all other pump photon energies, Fig.~\ref{fig: RT fit param}a shows that the fast decay time $\tau_1$ ranges from 100~fs to 300~fs, increasing with excitation fluence. This fluence-dependent slowing down is consistent with predictions from two-temperature model rate equations, in which stronger excitation produces a larger instantaneous rise in electronic temperature, thereby reducing the recovery rate~\cite{allen1987theory,giannetti2016ultrafast}. 

The slow decay time $\tau_2$ (Fig.~\ref{fig: RT fit param}b) is $\sim$2--5~ps, with large uncertainties preventing the identification of clear trends with either pump photon energy or excitation density. 

The amplitude parameters $a_1$, $a_2$, and $c$ increase approximately linearly with $I_{\mathrm{exc}}$ for all $E_{\mathrm{EUV}}$ values (see Fig. S4, Supplementary Material \cite{supplemental}). To compare amplitudes across pump photon energies, we need to account for variations in the fraction of the probed volume that is excited, $v = V_{\mathrm{exc}} / V_{\mathrm{probe}}$. For this reason, in Fig. \ref{fig: RT fit param}c--e we report the parameters $a_1$, $a_2$ and $c$ normalized by the normalization factor $r$ obtained from the transfer matrix method reflectivity model (see Supplementary Materials Sec. S2 \cite{supplemental}), while the not-normalized values are provided in Fig. S4. This normalization yields comparable amplitude values for all $E_{\mathrm{EUV}}$ at a given excitation density, indicating that differences in $\Delta R/R$ amplitude arise primarily from the short EUV absorption length and the resulting variation in excited volume.

These results demonstrate that the room-temperature response is largely independent of the EUV pump photon energy. This universality also extends to lower photon energies: in particular, we compare the EUV-pump–induced dynamics with those obtained under conventional optical excitation at 1.5~eV. The optical-pump trace (adapted from Ref. \citenum{giannetti2009discontinuity}), overlaid with the FEL-pump trace at $E_{\mathrm{EUV}} = 77$~eV in Fig.~\ref{fig: RT dynamics}b, is characterized by decay times $\tau_1 = 190 \pm 40$~fs and $\tau_2 = 1.3 \pm 0.8$~ps (see Fig. S7 Supplementary Material \cite{supplemental} for the complete set of fit parameters), in excellent agreement with the dynamics measured under FEL excitation.

In addition to the decay dynamics discussed above, the FEL-pump high-temperature data consistently reveal a weak oscillatory component in the $\sim3-20$~ps range. This oscillation has a period of approximately 6 ps, corresponding to a frequency of about 150 GHz, as extracted from fits reported in the Supplementary Material \cite{supplemental} (see Fig. S5 and S6 Supplementary Material). The oscillatory behavior is generally absent when the transient reflectivity signal is weak—either at low excitation fluence or when using the 23.5 eV pump. The 150 GHz observed frequency is too high to originate from coherent acoustic (Brillouin-type) phonons; in time-domain Brillouin scattering, the frequency is given by $f_B = 2n v_s / \lambda_{\text{probe}}$, where $n$ is the refractive index, $v_s$ the sound velocity and $\lambda_{\text{probe}}$ is the probe wavelength \cite{thomsen1986surface}. Considering $n = 1.5-1.6$ \cite{giannetti2011revealing}, $v_s = 2.5–5$~kms$^{-1}$ \cite{saunders1994anisotropy,wang1989elastic}, and $\lambda_{\text{probe}} = 785–830$~nm, one expects $f_B \approx 9–20$~GHz ($50-110$~ps), which are an order of magnitude lower frequency compared to the detected one, ruling out an acoustic origin. While the origin of such oscillation remains unclear, it could be compatible with an ultra-low-frequency Raman-active excitation, as similarly reported in Bi2212~\cite{mcniven2022long}. Interestingly, this feature is observed only under EUV excitation and not in conventional optical-pump measurements, although in the latter case comparable signal amplitudes may not be available.

\subsection{Superconducting phase response}

\begin{figure*}
\centering
\includegraphics[width=18cm]{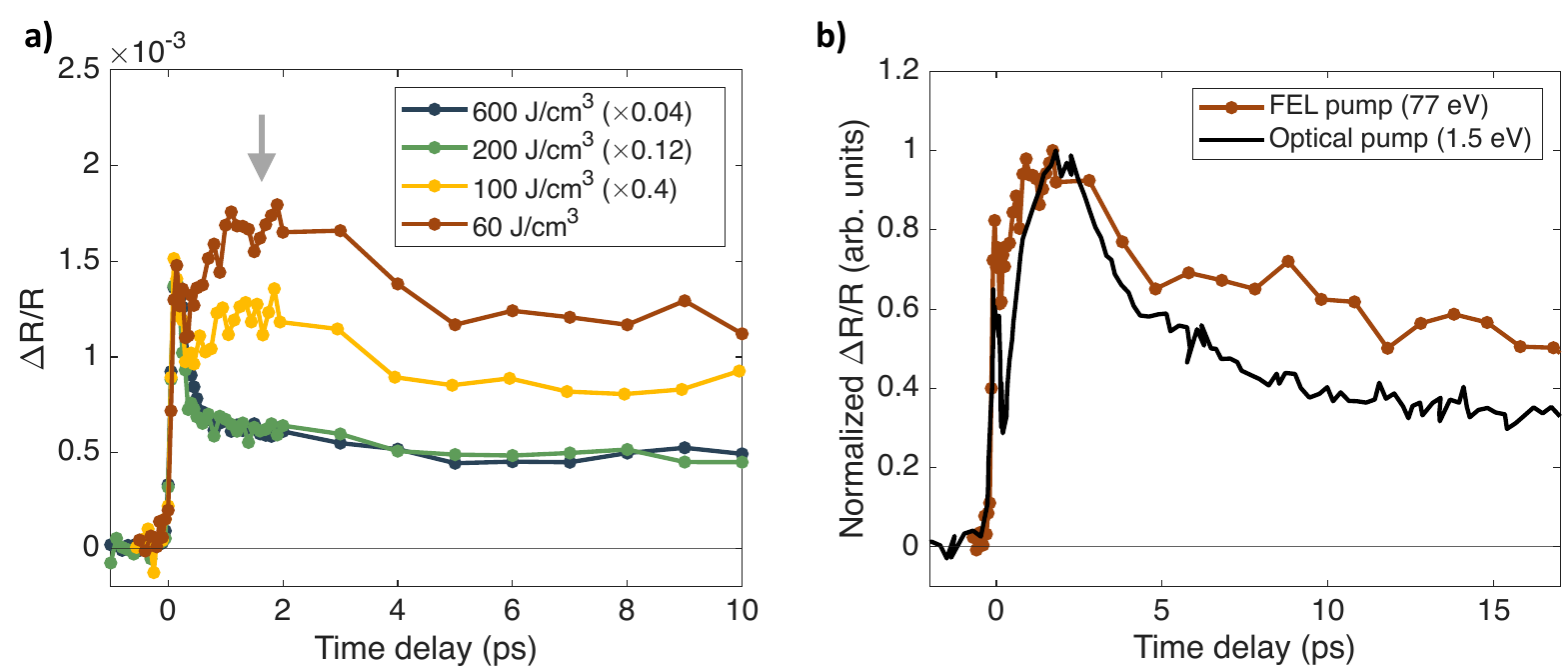}
\caption{Transient reflectivity dynamics in the superconducting state ($T < T_c$). 
(a) Fluence dependence of the $\Delta R/R$ response measured at $T = 30$~K with $E_{\mathrm{EUV}}=77$~eV for various excitation energy densities $I_{\mathrm{EUV}}$. 
High-fluence traces are rescaled by the normalization factors indicated in the legend to enable direct comparison of temporal profiles. 
For $I_{\mathrm{EUV}} < 100$~J/cm$^3$, an additional component, highlighted by the grey arrow, emerges with a build-up time of several hundred femtoseconds. 
This feature, absent in the high-fluence or high-temperature response, signals incomplete suppression of superconductivity. 
(b) Comparison of EUV-pump ($E_{\mathrm{EUV}} = 77$~eV) and near-IR-pump ($E_{\mathrm{pump}} = 1.5$~eV) dynamics. 
The FEL-pump trace was recorded at $T = 30$~K with $F = 220$~\textmu J/cm$^2$ ($I_{\mathrm{EUV}} = 60$~J/cm$^3$), 
while the optical-pump trace (adapted from Ref.~\citenum{giannetti2009discontinuity}) was measured at $T = 20$~K with $F = 285$~\textmu J/cm$^2$ ($I_{\mathrm{opt}} = 11$~J/cm$^3$).}
\label{fig: low T dynamics}
\end{figure*}

Low-temperature measurements were performed at $T = 30$ K ($< T_c$) with $E_{\mathrm{EUV}} = 77$ eV to investigate the temporal response to EUV excitation in the superconducting phase, assess whether superconductivity can be partially suppressed or fully quenched, and determine the corresponding critical fluence.

Cooling the sample below $T_c$ produces a striking change in the transient reflectivity response, whose temporal shape — rather than just its amplitude — becomes strongly dependent on the excitation fluence (Fig.~\ref{fig: low T dynamics}a).
At low excitation densities ($I_{\mathrm{EUV}} \lesssim 100$~J/cm$^3$), the optical dynamics deviate markedly from the room-temperature behavior: a delayed reflectivity component emerges (indicated by the gray arrow in Fig.~\ref{fig: low T dynamics}a), building up over several hundred femtoseconds and peaking at $\sim$1.5 ps at the lowest measured fluence. This slow component, absent in the normal state, is characteristic of quasiparticle recombination and condensate recovery, and is attributed to a boson bottleneck mechanism \cite{giannetti2009discontinuity,giannetti2016ultrafast}.
As the fluence increases ($I_{\mathrm{EUV}} =200–1000$~J/cm$^3$), this delayed component vanishes, and the signal evolves toward the normal-state profile. At fluences $I_{\mathrm{EUV}} \geq 200$ J/cm$^3$, the dynamics are dominated by the same fast decay (100–300 fs, depending on intensity) observed at room temperature. The disappearance of the slow build-up channel at large excitation intensity indicates complete suppression of the superconducting condensate under these conditions. This abrupt crossover underscores the role of excitation density in driving the system across the phase boundary and defines a threshold regime for EUV-induced condensate destruction.

The observed behavior closely mirrors previous optical-pump/optical-probe measurements on cuprates~\cite{giannetti2009discontinuity}, as well as additional near-IR results acquired for comparison (see Fig. S9 Supplementary Material \cite{supplemental}). In these cases, a delayed peak appears at excitation energy density $I_{\mathrm{opt}}$ up to $\sim$23~J/cm$^3$ (corresponding to fluence $F\approx600$~\textmu J/cm$^2$) and shifts to longer delays as the fluence increases (see Fig. S9b Supplementary Material \cite{supplemental}). Based on the data reported in Ref. \citenum{giannetti2009discontinuity}, for instance, with a near-IR pump excitation intensity of $\sim$10~J/cm$^3$ ($F\sim$250 $\mu$J/cm$^2$), the transient reflectivity dynamics resembles those measured with the FEL pump at the lowest EUV fluence (red trace in Fig. \ref{fig: low T dynamics}a), peaking at $\sim$1.5 ps.
Figures~\ref{fig: low T dynamics}b compares the low-temperature dynamics for the two excitation schemes: EUV pumping at 77~eV and near-IR pumping at 1.5~eV. The optical-pump trace, adapted from Ref.~\citenum{giannetti2009discontinuity}, was taken at $T = 20$ K and $F = 285$ $\mu$J/cm$^2$; the FEL-pump trace was recorded at $T = 30$ K and $F = 220$ $\mu$J/cm$^2$. Despite the similar nominal fluences, the corresponding absorbed energy densities differ substantially — $\sim$60 J/cm$^3$ for EUV versus $\sim$11 J/cm$^3$ for near-IR — due to the greater penetration depth of the near-IR light. This $\sim$6-fold discrepancy reflects the ratio of volumes excited with the near-IR and FEL pumps ($V_{\mathrm{exc}}^{\mathrm{opt}}/V_{\mathrm{exc}}^{\mathrm{FEL}} \approx 6.8$), explaining why higher absorbed energy density is required for EUV excitation to achieve dynamics comparable to those obtained with near-IR light.

\begin{figure*}
\centering
\includegraphics[width=18cm]{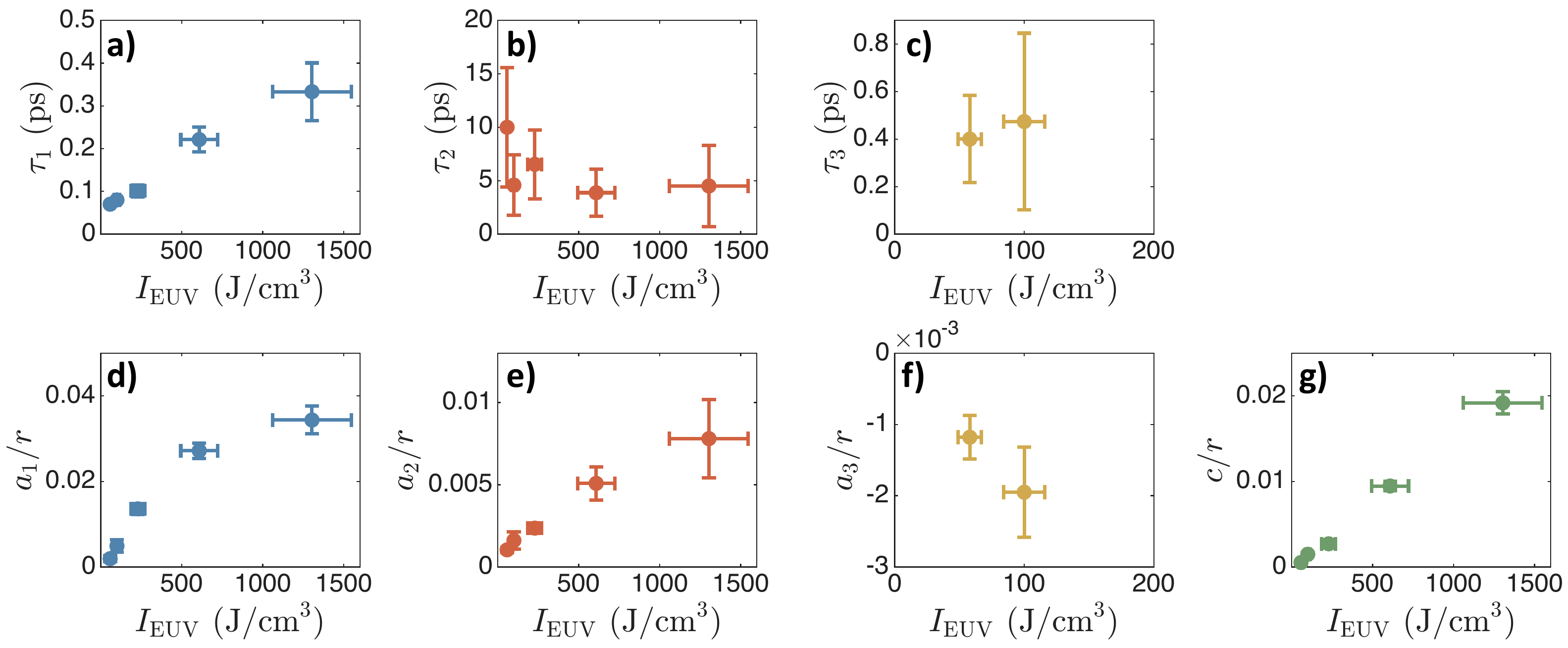}
\caption{Fit parameters extracted from low-temperature ($T < T_c$) transient reflectivity measurements as a function of excitation energy density $I_{\mathrm{EUV}}$.
a–c) Decay times $\tau_1$, $\tau_2$, and $\tau_3$, respectively.
d–g) Amplitude parameters $a_1$, $a_2$, $a_3$, and $c$, normalized to the parameter $r = 1.14$ accounting for the pump probe absorption length mismatch for consistency with Fig.~\ref{fig: RT fit param}. The $a_3$ term is included only for the two lowest measured excitation fluences, where the slow build-up component associated with superconducting condensate recovery is observed.
}
\label{fig: low T fit}
\end{figure*}

\begin{figure}
\centering
\includegraphics[width=8.5cm]{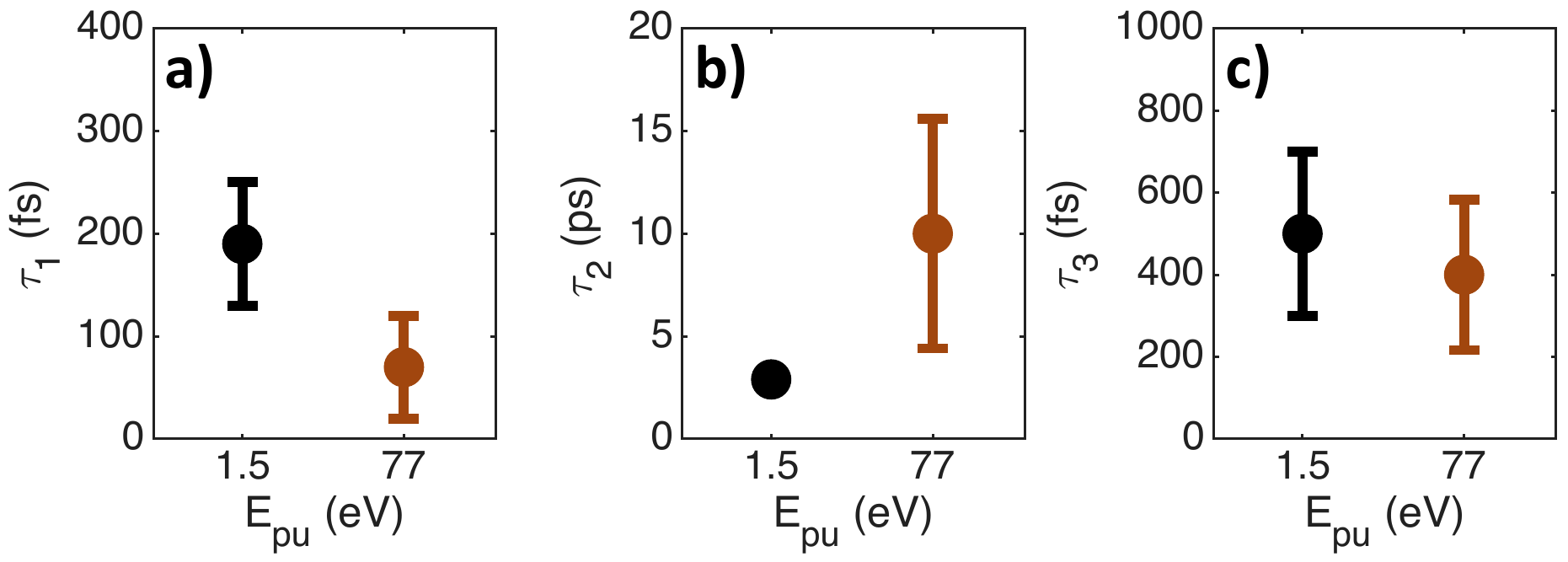}
\caption{Characteristic times of the transient reflectivity dynamics induced by near-IR 1.5~eV pump (black markers) and FEL 77~eV pump (red markers) at $T<T_c$: a) fast decay time $\tau_1$, b) slow decay time $\tau_2$, c) build-up time $\tau_3$.
}
\label{fig: low T fit param}
\end{figure}

$\Delta R/R$ dynamics fits for $T < T_c$ were performed using the expression
\begin{equation}
I_{\mathrm{LT}} (\Delta t) = a_1 e^{-\Delta t / \tau_1} + a_2 e^{-\Delta t / \tau_2} + a_3 e^{-\Delta t / \tau_3} + c,
\end{equation}
multiplied by a Heaviside step function and convoluted with a Gaussian of FWHM 120 fs. The resulting fit parameters are shown in Fig. \ref{fig: low T fit}, while the full set of fitted traces is reported in Fig. S8 Supplementary Materials \cite{supplemental}. For excitation intensities $\leq 100$ J/cm$^3$, an additional term with negative amplitude ($a_3$), compared to the room-temperature fit function, is required to capture the delayed “hump” associated with condensate recovery (Fig. \ref{fig: low T fit}c and f). At the lowest excitation intensity, we find $\tau_3 = 400 \pm 200$ fs. Applying the same fitting procedure to the optical-pump data in Fig. \ref{fig: low T dynamics}b yields $\tau_3 = 500 \pm 200$ fs (see Fig. S9a Supplementary Materials \cite{supplemental}). A comparison of the characteristic times $\tau_i$ obtained with 1.5~eV and 77~eV pump photon energies is shown in Fig.~\ref{fig: low T fit param}. The similar order of magnitude of $\tau_1$ and $\tau_2$, and compatible values of $\tau_3$, in the two excitation schemes confirms the comparable underlying dynamics, with small differences attributable to the different temperatures and excitation intensities in the respective measurements.
This finding demonstrates a scaling collapse of the relaxation dynamics when expressed as a function of the absorbed energy density $I_{\mathrm{exc}}$, confirming that energy density---not excitation pathway---governs the ultrafast electronic relaxation.

\section{Discussion: Possible Excitation Mechanism Above and Below $T_c$}

The striking similarity between the transient optical responses induced by EUV (77~eV) and near-infrared (1.5~eV) pulses—both above and below the superconducting transition temperature—suggests that the underlying excitation mechanism is governed more by the absorbed energy density than by the specific nature of the initial photoexcited states. While EUV photons nominally excite deep valence or even core-level electrons, the data indicate that the energy is rapidly redistributed through a cascade of secondary scattering events, leading to a thermalized population of high-energy quasiparticles in the conduction band. This ultrafast energy redistribution appears to funnel the initial excitation into the same low-energy degrees of freedom that are directly accessed via near-IR excitation.

Above $T_c$, both EUV and near-IR excitation generate a broad continuum of quasiparticles, likely originating from charge-transfer transitions involving O-2p and Cu-3d states. These carriers quickly relax into lower-lying electronic states via electron–electron and electron–phonon scattering. The measured reflectivity dynamics—featuring a fast rise and a multicomponent decay—suggest a rapid establishment of a quasi-thermal electronic distribution, regardless of whether it is seeded by a shallow interband transition or a deep photoionization event.

Below $T_c$, a distinct delayed component emerges in the reflectivity signal, peaking at $\sim$1.5~ps for both EUV and near-IR pump types at low fluence. This feature is a hallmark of quasiparticle recombination and superconducting gap recovery \cite{giannetti2009discontinuity}. Its persistence under EUV excitation indicates that the superconducting condensate is not immediately destroyed, but rather partially depleted and allowed to recover. This observation implies that, despite their high photon energy, EUV pulses can couple to the superconducting order, likely through their ability to generate dense quasiparticle populations that interact with the condensate indirectly. The recombination dynamics then follow similar bottleneck-limited pathways as seen in optical excitation, involving boson-mediated pair recovery processes.
Overall, these results support a picture in which the macroscopic ultrafast dynamics—such as quasiparticle relaxation and condensate recovery—are largely independent of the microscopic excitation pathway, provided the deposited energy exceeds the relevant thresholds. This universality suggests that even high-energy EUV photons, typically associated with core-level spectroscopy, can be harnessed to manipulate low-energy quantum states in a controlled manner, provided the excitation density is appropriately calibrated. Such insights open the door to extending control of correlated phases into new spectral and spatial regimes using FEL-based ultrafast techniques.
\section{Outlook: Speculative Dynamics under Soft X-ray and Attosecond Excitation in Cuprates}

Extending ultrafast spectroscopy of cuprates to the soft X-ray and attosecond regimes enables access to correlation dynamics on intrinsic timescales \cite{Chen2019trRIXS,Mukamel2013}. Unlike optical pumping, soft X-rays at the O-1s or Cu-2p edges create localized core holes that decay via Auger or radiative processes within a few femtoseconds, producing hot carriers and structural distortions. These local perturbations can suppress superconductivity non-thermally or disrupt pseudogap correlations on femtosecond scales \cite{Madan2014}.
Attosecond soft X-ray pulses ($<300$~as, $>200$~eV) allow direct tracking of core-hole creation, screening, and charge redistribution \cite{Sidiropoulos2021,Mukamel2013}. In cuprates, this could reveal valence charge transfer, gap renormalization, and doublon–holon dynamics. Experimental approaches include attosecond pump–probe, streaking, and two-color schemes to disentangle electronic and lattice responses.
Furthermore, attosecond soft X-ray experiments promise a revolutionary view into the microscopic choreography of electronic correlations. In cuprates, such tools could directly interrogate how Mott physics, charge transfer, and superconducting order dynamically emerge, collapse, or compete—all before lattice responses even begin.

FEL-based attosecond sources promise sufficient fluence for pump–probe control, while HHG-based sources remain powerful probes in the soft X-ray regime \cite{Seddon2017,Huang2021,Johnson2019,li2020attosecond,Huang2016}. Combined with advanced many-body simulations (tDMRG, non-equilibrium DMFT), these approaches could open new routes to probe and manipulate correlated quantum states.\\

\section{Conclusions}

Our experiment establishes a energy-density–driven pathway for electronic excitations, in clear contrast to bosonic driving where resonance-selective dynamics prevail. This underscores the need to distinguish electronic and bosonic routes to nonequilibrium superconductivity in the Bi2212 cuprate superconductor.\\
Deep UV or X-rays and optical photons excite very different states, yet both lead to the same quasiparticle bottlenecks and condensate recovery, highlighting a fundamental universality of electronic excitations in cuprates.\\

Using FEL-EUV pulses at 77~eV, we have shown that it is possible to perturb the superconducting state without immediately destroying it, even when the excitation originates from deep-lying electronic levels. Remarkably, the observed transient reflectivity dynamics closely mirror those triggered by conventional 1.5~eV near-IR pumping across both the superconducting and normal phases. This striking similarity suggests a universal photoresponse that is governed more by the absorbed energy density than by the specific nature of the initial excitation pathway.

Below the superconducting transition temperature, both excitation schemes give rise to a delayed reflectivity component indicative of quasiparticle recombination and condensate recovery. The emergence of this feature under EUV pumping demonstrates that the superconducting state retains its coherence even under high-energy excitation, as long as the energy density remains below the destruction threshold. The main difference lies in the amount of absorbed energy required: due to the much shorter EUV penetration depth, approximately an order of magnitude more energy must be deposited in a localized region to produce a comparable effect, leading to inhomogeneous excitation and a diluted probe signal.

Our experiment represents a first step toward clarifying whether the observed dynamics reflect direct coupling to the condensate or are instead mediated by indirect scattering cascades. This question could be further addressed by complementing our data with theoretical modeling, such as non-equilibrium dynamical mean-field theory (NE-DMFT) \cite{Georges1996,Eckstein2009,Amaricci2015,Werner2012} or time-dependent density matrix renormalization group (tDMRG) \cite{Schollwock2011,Kollath2007,Barmettler2008,Daley2004} studies focused on cascade thermalization or multi-step relaxation, where an initial non-thermal distribution of high-energy excitations (e.g., electrons coupled to bosons) relaxes via intermediate scattering channels, before reaching thermal equilibrium.\\
The present results lay the foundation for future ultrafast studies of correlated superconductors by identifying the experimental conditions under which out-of-equilibrium dynamics reflect intrinsic electronic correlations rather than the specifics of the optical coupling channel. We find that in Bi2212 the transient electronic dynamics are governed solely by the absorbed energy density, even when excitation pathways range from interband transitions to deep-core photoionization. This universality provides a benchmark for future FEL and attosecond investigations of correlated materials and establishes FEL-based EUV excitations as a foundation for next-generation pump–probe studies of quantum materials, enabling access to regimes beyond optical techniques and opening pathways toward soft X-ray and attosecond multidimensional spectroscopies of correlated phases.

\begin{acknowledgments}
We gratefully acknowledge the outstanding professionalism and continuous support of the FERMI FEL team and the beamline optics group, whose expertise in operating the facility and handling the transport optics of the FEL pulses was essential to the success of this work.\\
The work at BNL was supported by the US Department of Energy, office of Basic Energy Sciences, contract no. DOE-SC0012704.\\
A.M., Francesco P. and C.G. acknowledge financial support from MIUR through the PRIN 2020 (Prot. 2020JLZ52N 003) program and from the European Union - Next Generation EU through the MUR-PRIN2022 (Prot. 20228YCYY7) program.

\end{acknowledgments}
%Furthermore, attosecond soft X-ray experiments promise a revolutionary view into the microscopic choreography of electronic correlations. In cuprates, such tools could directly interrogate how Mott physics, charge transfer, and superconducting order dynamically emerge, collapse, or compete—all before lattice responses even begin. These experiments would push the boundaries of both temporal and spatial resolution in strongly correlated materials and non-conventional superconductor

\section*{Data Availability}
The data are available from the authors upon reasonable request.
\bibliography{Refs}

\begin{thebibliography}{2}%
\makeatletter
\providecommand \@ifxundefined [1]{%
 \@ifx{#1\undefined}
}%
\providecommand \@ifnum [1]{%
 \ifnum #1\expandafter \@firstoftwo
 \else \expandafter \@secondoftwo
 \fi
}%
\providecommand \@ifx [1]{%
 \ifx #1\expandafter \@firstoftwo
 \else \expandafter \@secondoftwo
 \fi
}%
\providecommand \natexlab [1]{#1}%
\providecommand \enquote  [1]{``#1''}%
\providecommand \bibnamefont  [1]{#1}%
\providecommand \bibfnamefont [1]{#1}%
\providecommand \citenamefont [1]{#1}%
\providecommand \href@noop [0]{\@secondoftwo}%
\providecommand \href [0]{\begingroup \@sanitize@url \@href}%
\providecommand \@href[1]{\@@startlink{#1}\@@href}%
\providecommand \@@href[1]{\endgroup#1\@@endlink}%
\providecommand \@sanitize@url [0]{\catcode `\\12\catcode `\$12\catcode
  `\&12\catcode `\#12\catcode `\^12\catcode `\_12\catcode `\%12\relax}%
\providecommand \@@startlink[1]{}%
\providecommand \@@endlink[0]{}%
\providecommand \url  [0]{\begingroup\@sanitize@url \@url }%
\providecommand \@url [1]{\endgroup\@href {#1}{\urlprefix }}%
\providecommand \urlprefix  [0]{URL }%
\providecommand \Eprint [0]{\href }%
\providecommand \doibase [0]{https://doi.org/}%
\providecommand \selectlanguage [0]{\@gobble}%
\providecommand \bibinfo  [0]{\@secondoftwo}%
\providecommand \bibfield  [0]{\@secondoftwo}%
\providecommand \translation [1]{[#1]}%
\providecommand \BibitemOpen [0]{}%
\providecommand \bibitemStop [0]{}%
\providecommand \bibitemNoStop [0]{.\EOS\space}%
\providecommand \EOS [0]{\spacefactor3000\relax}%
\providecommand \BibitemShut  [1]{\csname bibitem#1\endcsname}%
\let\auto@bib@innerbib\@empty
%</preamble>
\bibitem [{\citenamefont {Giannetti}\ \emph {et~al.}(2011)\citenamefont
  {Giannetti}, \citenamefont {Cilento}, \citenamefont {Conte}, \citenamefont
  {Coslovich}, \citenamefont {Ferrini}, \citenamefont {Molegraaf},
  \citenamefont {Raichle}, \citenamefont {Liang}, \citenamefont {Eisaki},
  \citenamefont {Greven} \emph {et~al.}}]{giannetti2011revealing}%
  \BibitemOpen
  \bibfield  {author} {\bibinfo {author} {\bibfnamefont {C.}~\bibnamefont
  {Giannetti}}, \bibinfo {author} {\bibfnamefont {F.}~\bibnamefont {Cilento}},
  \bibinfo {author} {\bibfnamefont {S.~D.}\ \bibnamefont {Conte}}, \bibinfo
  {author} {\bibfnamefont {G.}~\bibnamefont {Coslovich}}, \bibinfo {author}
  {\bibfnamefont {G.}~\bibnamefont {Ferrini}}, \bibinfo {author} {\bibfnamefont
  {H.}~\bibnamefont {Molegraaf}}, \bibinfo {author} {\bibfnamefont
  {M.}~\bibnamefont {Raichle}}, \bibinfo {author} {\bibfnamefont
  {R.}~\bibnamefont {Liang}}, \bibinfo {author} {\bibfnamefont
  {H.}~\bibnamefont {Eisaki}}, \bibinfo {author} {\bibfnamefont
  {M.}~\bibnamefont {Greven}}, \emph {et~al.},\ }\bibfield  {title} {\bibinfo
  {title} {Revealing the high-energy electronic excitations underlying the
  onset of high-temperature superconductivity in cuprates},\ }\href@noop {}
  {\bibfield  {journal} {\bibinfo  {journal} {Nat. Comm.}\ }\textbf {\bibinfo
  {volume} {2}},\ \bibinfo {pages} {353} (\bibinfo {year} {2011})}\BibitemShut
  {NoStop}%
\bibitem [{\citenamefont {Giannetti}\ \emph {et~al.}(2009)\citenamefont
  {Giannetti}, \citenamefont {Coslovich}, \citenamefont {Cilento},
  \citenamefont {Ferrini}, \citenamefont {Eisaki}, \citenamefont {Kaneko},
  \citenamefont {Greven},\ and\ \citenamefont
  {Parmigiani}}]{giannetti2009discontinuity}%
  \BibitemOpen
  \bibfield  {author} {\bibinfo {author} {\bibfnamefont {C.}~\bibnamefont
  {Giannetti}}, \bibinfo {author} {\bibfnamefont {G.}~\bibnamefont
  {Coslovich}}, \bibinfo {author} {\bibfnamefont {F.}~\bibnamefont {Cilento}},
  \bibinfo {author} {\bibfnamefont {G.}~\bibnamefont {Ferrini}}, \bibinfo
  {author} {\bibfnamefont {H.}~\bibnamefont {Eisaki}}, \bibinfo {author}
  {\bibfnamefont {N.}~\bibnamefont {Kaneko}}, \bibinfo {author} {\bibfnamefont
  {M.}~\bibnamefont {Greven}},\ and\ \bibinfo {author} {\bibfnamefont
  {F.}~\bibnamefont {Parmigiani}},\ }\bibfield  {title} {\bibinfo {title}
  {{Discontinuity of the ultrafast electronic response of underdoped
  superconducting
  ${\text{Bi}}_{2}{\text{Sr}}_{2}{\text{CaCu}}_{2}{\text{O}}_{8+\ensuremath{\delta}}$
  strongly excited by ultrashort light pulses}},\ }\href
  {https://doi.org/10.1103/PhysRevB.79.224502} {\bibfield  {journal} {\bibinfo
  {journal} {Phys. Rev. B}\ }\textbf {\bibinfo {volume} {79}},\ \bibinfo
  {pages} {224502} (\bibinfo {year} {2009})}\BibitemShut {NoStop}%
\end{thebibliography}%


\begin{thebibliography}{54}%
\makeatletter
\providecommand \@ifxundefined [1]{%
 \@ifx{#1\undefined}
}%
\providecommand \@ifnum [1]{%
 \ifnum #1\expandafter \@firstoftwo
 \else \expandafter \@secondoftwo
 \fi
}%
\providecommand \@ifx [1]{%
 \ifx #1\expandafter \@firstoftwo
 \else \expandafter \@secondoftwo
 \fi
}%
\providecommand \natexlab [1]{#1}%
\providecommand \enquote  [1]{``#1''}%
\providecommand \bibnamefont  [1]{#1}%
\providecommand \bibfnamefont [1]{#1}%
\providecommand \citenamefont [1]{#1}%
\providecommand \href@noop [0]{\@secondoftwo}%
\providecommand \href [0]{\begingroup \@sanitize@url \@href}%
\providecommand \@href[1]{\@@startlink{#1}\@@href}%
\providecommand \@@href[1]{\endgroup#1\@@endlink}%
\providecommand \@sanitize@url [0]{\catcode `\\12\catcode `\$12\catcode
  `\&12\catcode `\#12\catcode `\^12\catcode `\_12\catcode `\%12\relax}%
\providecommand \@@startlink[1]{}%
\providecommand \@@endlink[0]{}%
\providecommand \url  [0]{\begingroup\@sanitize@url \@url }%
\providecommand \@url [1]{\endgroup\@href {#1}{\urlprefix }}%
\providecommand \urlprefix  [0]{URL }%
\providecommand \Eprint [0]{\href }%
\providecommand \doibase [0]{https://doi.org/}%
\providecommand \selectlanguage [0]{\@gobble}%
\providecommand \bibinfo  [0]{\@secondoftwo}%
\providecommand \bibfield  [0]{\@secondoftwo}%
\providecommand \translation [1]{[#1]}%
\providecommand \BibitemOpen [0]{}%
\providecommand \bibitemStop [0]{}%
\providecommand \bibitemNoStop [0]{.\EOS\space}%
\providecommand \EOS [0]{\spacefactor3000\relax}%
\providecommand \BibitemShut  [1]{\csname bibitem#1\endcsname}%
\let\auto@bib@innerbib\@empty
%</preamble>
\bibitem [{\citenamefont {Giannetti}\ \emph {et~al.}(2016)\citenamefont
  {Giannetti}, \citenamefont {Capone}, \citenamefont {Fausti}, \citenamefont
  {Fabrizio}, \citenamefont {Parmigiani},\ and\ \citenamefont
  {Mihailovic}}]{giannetti2016ultrafast}%
  \BibitemOpen
  \bibfield  {author} {\bibinfo {author} {\bibfnamefont {C.}~\bibnamefont
  {Giannetti}}, \bibinfo {author} {\bibfnamefont {M.}~\bibnamefont {Capone}},
  \bibinfo {author} {\bibfnamefont {D.}~\bibnamefont {Fausti}}, \bibinfo
  {author} {\bibfnamefont {M.}~\bibnamefont {Fabrizio}}, \bibinfo {author}
  {\bibfnamefont {F.}~\bibnamefont {Parmigiani}},\ and\ \bibinfo {author}
  {\bibfnamefont {D.}~\bibnamefont {Mihailovic}},\ }\bibfield  {title}
  {\bibinfo {title} {Ultrafast optical spectroscopy of strongly correlated
  materials and high-temperature superconductors: a non-equilibrium approach},\
  }\href@noop {} {\bibfield  {journal} {\bibinfo  {journal} {Advances in
  Physics}\ }\textbf {\bibinfo {volume} {65}},\ \bibinfo {pages} {58} (\bibinfo
  {year} {2016})}\BibitemShut {NoStop}%
\bibitem [{\citenamefont {Keimer}\ \emph {et~al.}(2015)\citenamefont {Keimer},
  \citenamefont {Kivelson}, \citenamefont {Norman}, \citenamefont {Uchida},\
  and\ \citenamefont {Zaanen}}]{keimer2015quantum}%
  \BibitemOpen
  \bibfield  {author} {\bibinfo {author} {\bibfnamefont {B.}~\bibnamefont
  {Keimer}}, \bibinfo {author} {\bibfnamefont {S.~A.}\ \bibnamefont
  {Kivelson}}, \bibinfo {author} {\bibfnamefont {M.~R.}\ \bibnamefont
  {Norman}}, \bibinfo {author} {\bibfnamefont {S.}~\bibnamefont {Uchida}},\
  and\ \bibinfo {author} {\bibfnamefont {J.}~\bibnamefont {Zaanen}},\
  }\bibfield  {title} {\bibinfo {title} {From quantum matter to
  high-temperature superconductivity in copper oxides},\ }\href@noop {}
  {\bibfield  {journal} {\bibinfo  {journal} {Nature}\ }\textbf {\bibinfo
  {volume} {518}},\ \bibinfo {pages} {179} (\bibinfo {year}
  {2015})}\BibitemShut {NoStop}%
\bibitem [{\citenamefont {Matsunaga}\ \emph {et~al.}(2014)\citenamefont
  {Matsunaga}, \citenamefont {Tsuji}, \citenamefont {Fujita}, \citenamefont
  {Sugioka}, \citenamefont {Makise}, \citenamefont {Uzawa}, \citenamefont
  {Terai}, \citenamefont {Wang}, \citenamefont {Aoki},\ and\ \citenamefont
  {Shimano}}]{matsunaga2014light}%
  \BibitemOpen
  \bibfield  {author} {\bibinfo {author} {\bibfnamefont {R.}~\bibnamefont
  {Matsunaga}}, \bibinfo {author} {\bibfnamefont {N.}~\bibnamefont {Tsuji}},
  \bibinfo {author} {\bibfnamefont {H.}~\bibnamefont {Fujita}}, \bibinfo
  {author} {\bibfnamefont {A.}~\bibnamefont {Sugioka}}, \bibinfo {author}
  {\bibfnamefont {K.}~\bibnamefont {Makise}}, \bibinfo {author} {\bibfnamefont
  {Y.}~\bibnamefont {Uzawa}}, \bibinfo {author} {\bibfnamefont
  {H.}~\bibnamefont {Terai}}, \bibinfo {author} {\bibfnamefont
  {Z.}~\bibnamefont {Wang}}, \bibinfo {author} {\bibfnamefont {H.}~\bibnamefont
  {Aoki}},\ and\ \bibinfo {author} {\bibfnamefont {R.}~\bibnamefont
  {Shimano}},\ }\bibfield  {title} {\bibinfo {title} {{Light-induced collective
  pseudospin precession resonating with Higgs mode in a superconductor}},\
  }\href@noop {} {\bibfield  {journal} {\bibinfo  {journal} {Science}\ }\textbf
  {\bibinfo {volume} {345}},\ \bibinfo {pages} {1145} (\bibinfo {year}
  {2014})}\BibitemShut {NoStop}%
\bibitem [{\citenamefont {Hu}\ \emph {et~al.}(2014)\citenamefont {Hu},
  \citenamefont {Kaiser}, \citenamefont {Nicoletti}, \citenamefont {Hunt},
  \citenamefont {Gierz}, \citenamefont {Hoffmann}, \citenamefont {Le~Tacon},
  \citenamefont {Loew}, \citenamefont {Keimer},\ and\ \citenamefont
  {Cavalleri}}]{hu2014optically}%
  \BibitemOpen
  \bibfield  {author} {\bibinfo {author} {\bibfnamefont {W.}~\bibnamefont
  {Hu}}, \bibinfo {author} {\bibfnamefont {S.}~\bibnamefont {Kaiser}}, \bibinfo
  {author} {\bibfnamefont {D.}~\bibnamefont {Nicoletti}}, \bibinfo {author}
  {\bibfnamefont {C.~R.}\ \bibnamefont {Hunt}}, \bibinfo {author}
  {\bibfnamefont {I.}~\bibnamefont {Gierz}}, \bibinfo {author} {\bibfnamefont
  {M.~C.}\ \bibnamefont {Hoffmann}}, \bibinfo {author} {\bibfnamefont
  {M.}~\bibnamefont {Le~Tacon}}, \bibinfo {author} {\bibfnamefont
  {T.}~\bibnamefont {Loew}}, \bibinfo {author} {\bibfnamefont {B.}~\bibnamefont
  {Keimer}},\ and\ \bibinfo {author} {\bibfnamefont {A.}~\bibnamefont
  {Cavalleri}},\ }\bibfield  {title} {\bibinfo {title} {{Optically enhanced
  coherent transport in \ch{YBa2Cu3O_{6.5}} by ultrafast redistribution of
  interlayer coupling}},\ }\href@noop {} {\bibfield  {journal} {\bibinfo
  {journal} {Nat. Mat.}\ }\textbf {\bibinfo {volume} {13}},\ \bibinfo {pages}
  {705} (\bibinfo {year} {2014})}\BibitemShut {NoStop}%
\bibitem [{\citenamefont {Liu}\ \emph {et~al.}(2020)\citenamefont {Liu},
  \citenamefont {F{\"o}rst}, \citenamefont {Fechner}, \citenamefont
  {Nicoletti}, \citenamefont {Porras}, \citenamefont {Loew}, \citenamefont
  {Keimer},\ and\ \citenamefont {Cavalleri}}]{liu2020pump}%
  \BibitemOpen
  \bibfield  {author} {\bibinfo {author} {\bibfnamefont {B.}~\bibnamefont
  {Liu}}, \bibinfo {author} {\bibfnamefont {M.}~\bibnamefont {F{\"o}rst}},
  \bibinfo {author} {\bibfnamefont {M.}~\bibnamefont {Fechner}}, \bibinfo
  {author} {\bibfnamefont {D.}~\bibnamefont {Nicoletti}}, \bibinfo {author}
  {\bibfnamefont {J.}~\bibnamefont {Porras}}, \bibinfo {author} {\bibfnamefont
  {T.}~\bibnamefont {Loew}}, \bibinfo {author} {\bibfnamefont {B.}~\bibnamefont
  {Keimer}},\ and\ \bibinfo {author} {\bibfnamefont {A.}~\bibnamefont
  {Cavalleri}},\ }\bibfield  {title} {\bibinfo {title} {Pump frequency
  resonances for light-induced incipient superconductivity in \ch{YBa 2 Cu 3
  O_{6.5}}},\ }\href@noop {} {\bibfield  {journal} {\bibinfo  {journal}
  {Physical Review X}\ }\textbf {\bibinfo {volume} {10}},\ \bibinfo {pages}
  {011053} (\bibinfo {year} {2020})}\BibitemShut {NoStop}%
\bibitem [{\citenamefont {Katsumi}\ \emph {et~al.}(2018)\citenamefont
  {Katsumi}, \citenamefont {Tsuji}, \citenamefont {Hamada}, \citenamefont
  {Matsunaga}, \citenamefont {Schneeloch}, \citenamefont {Zhong}, \citenamefont
  {Gu}, \citenamefont {Aoki}, \citenamefont {Gallais},\ and\ \citenamefont
  {Shimano}}]{katsumi2018higgs}%
  \BibitemOpen
  \bibfield  {author} {\bibinfo {author} {\bibfnamefont {K.}~\bibnamefont
  {Katsumi}}, \bibinfo {author} {\bibfnamefont {N.}~\bibnamefont {Tsuji}},
  \bibinfo {author} {\bibfnamefont {Y.~I.}\ \bibnamefont {Hamada}}, \bibinfo
  {author} {\bibfnamefont {R.}~\bibnamefont {Matsunaga}}, \bibinfo {author}
  {\bibfnamefont {J.}~\bibnamefont {Schneeloch}}, \bibinfo {author}
  {\bibfnamefont {R.~D.}\ \bibnamefont {Zhong}}, \bibinfo {author}
  {\bibfnamefont {G.~D.}\ \bibnamefont {Gu}}, \bibinfo {author} {\bibfnamefont
  {H.}~\bibnamefont {Aoki}}, \bibinfo {author} {\bibfnamefont {Y.}~\bibnamefont
  {Gallais}},\ and\ \bibinfo {author} {\bibfnamefont {R.}~\bibnamefont
  {Shimano}},\ }\bibfield  {title} {\bibinfo {title} {Higgs mode in the d-wave
  superconductor \ch{Bi2Sr2CaCu2O_{8+ x}} driven by an intense terahertz
  pulse},\ }\href@noop {} {\bibfield  {journal} {\bibinfo  {journal} {Phys.
  Rev. Lett.}\ }\textbf {\bibinfo {volume} {120}},\ \bibinfo {pages} {117001}
  (\bibinfo {year} {2018})}\BibitemShut {NoStop}%
\bibitem [{\citenamefont {Kaiser}\ \emph {et~al.}(2014)\citenamefont {Kaiser},
  \citenamefont {Hunt}, \citenamefont {Nicoletti}, \citenamefont {Hu},
  \citenamefont {Gierz}, \citenamefont {Liu}, \citenamefont {Le~Tacon},
  \citenamefont {Loew}, \citenamefont {Haug}, \citenamefont {Keimer} \emph
  {et~al.}}]{Kaiser2012}%
  \BibitemOpen
  \bibfield  {author} {\bibinfo {author} {\bibfnamefont {S.}~\bibnamefont
  {Kaiser}}, \bibinfo {author} {\bibfnamefont {C.~R.}\ \bibnamefont {Hunt}},
  \bibinfo {author} {\bibfnamefont {D.}~\bibnamefont {Nicoletti}}, \bibinfo
  {author} {\bibfnamefont {W.}~\bibnamefont {Hu}}, \bibinfo {author}
  {\bibfnamefont {I.}~\bibnamefont {Gierz}}, \bibinfo {author} {\bibfnamefont
  {H.}~\bibnamefont {Liu}}, \bibinfo {author} {\bibfnamefont {M.}~\bibnamefont
  {Le~Tacon}}, \bibinfo {author} {\bibfnamefont {T.}~\bibnamefont {Loew}},
  \bibinfo {author} {\bibfnamefont {D.}~\bibnamefont {Haug}}, \bibinfo {author}
  {\bibfnamefont {B.}~\bibnamefont {Keimer}}, \emph {et~al.},\ }\bibfield
  {title} {\bibinfo {title} {{Optically induced coherent transport far above
  $T_c$ in underdoped YBa$_2$Cu$_3$O$_{6+x}$}},\ }\href@noop {} {\bibfield
  {journal} {\bibinfo  {journal} {Phys. Rev. B}\ }\textbf {\bibinfo {volume}
  {89}},\ \bibinfo {pages} {184516} (\bibinfo {year} {2014})}\BibitemShut
  {NoStop}%
\bibitem [{\citenamefont {Budden}\ \emph {et~al.}(2021)\citenamefont {Budden},
  \citenamefont {Gebert}, \citenamefont {Nicholson}, \citenamefont {Narang},
  \citenamefont {Chan}, \citenamefont {Hildebrand}, \citenamefont {Freiwald},
  \citenamefont {Mark}, \citenamefont {Sato}, \citenamefont {Cavalleri},\ and\
  \citenamefont {Ernstorfer}}]{Budden2021}%
  \BibitemOpen
  \bibfield  {author} {\bibinfo {author} {\bibfnamefont {M.}~\bibnamefont
  {Budden}}, \bibinfo {author} {\bibfnamefont {T.}~\bibnamefont {Gebert}},
  \bibinfo {author} {\bibfnamefont {C.~W.}\ \bibnamefont {Nicholson}}, \bibinfo
  {author} {\bibfnamefont {P.}~\bibnamefont {Narang}}, \bibinfo {author}
  {\bibfnamefont {C.~K.}\ \bibnamefont {Chan}}, \bibinfo {author}
  {\bibfnamefont {B.}~\bibnamefont {Hildebrand}}, \bibinfo {author}
  {\bibfnamefont {M.}~\bibnamefont {Freiwald}}, \bibinfo {author}
  {\bibfnamefont {C.}~\bibnamefont {Mark}}, \bibinfo {author} {\bibfnamefont
  {S.~A.}\ \bibnamefont {Sato}}, \bibinfo {author} {\bibfnamefont
  {A.}~\bibnamefont {Cavalleri}},\ and\ \bibinfo {author} {\bibfnamefont
  {R.}~\bibnamefont {Ernstorfer}},\ }\bibfield  {title} {\bibinfo {title}
  {Evidence for metastable photo-induced superconductivity in cuprates via
  intense far-infrared pulses},\ }\href
  {https://doi.org/10.1038/s41567-020-01148-1} {\bibfield  {journal} {\bibinfo
  {journal} {Nat. Phys.}\ }\textbf {\bibinfo {volume} {17}},\ \bibinfo {pages}
  {611} (\bibinfo {year} {2021})}\BibitemShut {NoStop}%
\bibitem [{\citenamefont {Casandruc}\ \emph {et~al.}(2015)\citenamefont
  {Casandruc}, \citenamefont {Nicoletti}, \citenamefont {Rajasekaran},
  \citenamefont {Laplace}, \citenamefont {Khanna}, \citenamefont {Gu},
  \citenamefont {Hill},\ and\ \citenamefont {Cavalleri}}]{Casandruc2015}%
  \BibitemOpen
  \bibfield  {author} {\bibinfo {author} {\bibfnamefont {E.}~\bibnamefont
  {Casandruc}}, \bibinfo {author} {\bibfnamefont {D.}~\bibnamefont
  {Nicoletti}}, \bibinfo {author} {\bibfnamefont {S.}~\bibnamefont
  {Rajasekaran}}, \bibinfo {author} {\bibfnamefont {Y.}~\bibnamefont
  {Laplace}}, \bibinfo {author} {\bibfnamefont {V.}~\bibnamefont {Khanna}},
  \bibinfo {author} {\bibfnamefont {G.}~\bibnamefont {Gu}}, \bibinfo {author}
  {\bibfnamefont {J.}~\bibnamefont {Hill}},\ and\ \bibinfo {author}
  {\bibfnamefont {A.}~\bibnamefont {Cavalleri}},\ }\bibfield  {title} {\bibinfo
  {title} {Wavelength-dependent optical enhancement of superconducting
  interlayer coupling in \ch{La_{1.885} Ba_{0.115} CuO_4}},\ }\href@noop {}
  {\bibfield  {journal} {\bibinfo  {journal} {Physical Review B}\ }\textbf
  {\bibinfo {volume} {91}},\ \bibinfo {pages} {174502} (\bibinfo {year}
  {2015})}\BibitemShut {NoStop}%
\bibitem [{\citenamefont {F{\"o}rst}\ \emph {et~al.}(2011)\citenamefont
  {F{\"o}rst}, \citenamefont {Manzoni}, \citenamefont {Kaiser}, \citenamefont
  {Tomioka}, \citenamefont {Tokura}, \citenamefont {Merlin},\ and\
  \citenamefont {Cavalleri}}]{forst2011nonlinear}%
  \BibitemOpen
  \bibfield  {author} {\bibinfo {author} {\bibfnamefont {M.}~\bibnamefont
  {F{\"o}rst}}, \bibinfo {author} {\bibfnamefont {C.}~\bibnamefont {Manzoni}},
  \bibinfo {author} {\bibfnamefont {S.}~\bibnamefont {Kaiser}}, \bibinfo
  {author} {\bibfnamefont {Y.}~\bibnamefont {Tomioka}}, \bibinfo {author}
  {\bibfnamefont {Y.}~\bibnamefont {Tokura}}, \bibinfo {author} {\bibfnamefont
  {R.}~\bibnamefont {Merlin}},\ and\ \bibinfo {author} {\bibfnamefont
  {A.}~\bibnamefont {Cavalleri}},\ }\bibfield  {title} {\bibinfo {title}
  {Nonlinear phononics as an ultrafast route to lattice control},\ }\href@noop
  {} {\bibfield  {journal} {\bibinfo  {journal} {Nat. Phys.}\ }\textbf
  {\bibinfo {volume} {7}},\ \bibinfo {pages} {854} (\bibinfo {year}
  {2011})}\BibitemShut {NoStop}%
\bibitem [{\citenamefont {Mankowsky}\ \emph {et~al.}(2014)\citenamefont
  {Mankowsky}, \citenamefont {Subedi}, \citenamefont {F{\"o}rst}, \citenamefont
  {Mariager}, \citenamefont {Chollet}, \citenamefont {Lemke}, \citenamefont
  {Robinson}, \citenamefont {Glownia}, \citenamefont {Minitti}, \citenamefont
  {Frano} \emph {et~al.}}]{mankowsky2014nonlinear}%
  \BibitemOpen
  \bibfield  {author} {\bibinfo {author} {\bibfnamefont {R.}~\bibnamefont
  {Mankowsky}}, \bibinfo {author} {\bibfnamefont {A.}~\bibnamefont {Subedi}},
  \bibinfo {author} {\bibfnamefont {M.}~\bibnamefont {F{\"o}rst}}, \bibinfo
  {author} {\bibfnamefont {S.~O.}\ \bibnamefont {Mariager}}, \bibinfo {author}
  {\bibfnamefont {M.}~\bibnamefont {Chollet}}, \bibinfo {author} {\bibfnamefont
  {H.}~\bibnamefont {Lemke}}, \bibinfo {author} {\bibfnamefont {J.~S.}\
  \bibnamefont {Robinson}}, \bibinfo {author} {\bibfnamefont {J.~M.}\
  \bibnamefont {Glownia}}, \bibinfo {author} {\bibfnamefont {M.~P.}\
  \bibnamefont {Minitti}}, \bibinfo {author} {\bibfnamefont {A.}~\bibnamefont
  {Frano}}, \emph {et~al.},\ }\bibfield  {title} {\bibinfo {title} {Nonlinear
  lattice dynamics as a basis for enhanced superconductivity in
  \ch{YBa2Cu3O_{6.5}}},\ }\href@noop {} {\bibfield  {journal} {\bibinfo
  {journal} {Nature}\ }\textbf {\bibinfo {volume} {516}},\ \bibinfo {pages}
  {71} (\bibinfo {year} {2014})}\BibitemShut {NoStop}%
\bibitem [{\citenamefont {Fausti}\ \emph {et~al.}(2011)\citenamefont {Fausti},
  \citenamefont {Tobey}, \citenamefont {Dean}, \citenamefont {Kaiser},
  \citenamefont {Dienst}, \citenamefont {Hoffmann}, \citenamefont {Pyon},
  \citenamefont {Takayama}, \citenamefont {Takagi},\ and\ \citenamefont
  {Cavalleri}}]{fausti2011light}%
  \BibitemOpen
  \bibfield  {author} {\bibinfo {author} {\bibfnamefont {D.}~\bibnamefont
  {Fausti}}, \bibinfo {author} {\bibfnamefont {R.}~\bibnamefont {Tobey}},
  \bibinfo {author} {\bibfnamefont {N.}~\bibnamefont {Dean}}, \bibinfo {author}
  {\bibfnamefont {S.}~\bibnamefont {Kaiser}}, \bibinfo {author} {\bibfnamefont
  {A.}~\bibnamefont {Dienst}}, \bibinfo {author} {\bibfnamefont {M.~C.}\
  \bibnamefont {Hoffmann}}, \bibinfo {author} {\bibfnamefont {S.}~\bibnamefont
  {Pyon}}, \bibinfo {author} {\bibfnamefont {T.}~\bibnamefont {Takayama}},
  \bibinfo {author} {\bibfnamefont {H.}~\bibnamefont {Takagi}},\ and\ \bibinfo
  {author} {\bibfnamefont {A.}~\bibnamefont {Cavalleri}},\ }\bibfield  {title}
  {\bibinfo {title} {Light-induced superconductivity in a stripe-ordered
  cuprate},\ }\href@noop {} {\bibfield  {journal} {\bibinfo  {journal}
  {Science}\ }\textbf {\bibinfo {volume} {331}},\ \bibinfo {pages} {189}
  (\bibinfo {year} {2011})}\BibitemShut {NoStop}%
\bibitem [{\citenamefont {Smallwood}\ \emph {et~al.}(2012)\citenamefont
  {Smallwood}, \citenamefont {Hinton}, \citenamefont {Jozwiak}, \citenamefont
  {Zhang}, \citenamefont {Koralek}, \citenamefont {Eisaki}, \citenamefont
  {Lee}, \citenamefont {Orenstein},\ and\ \citenamefont
  {Lanzara}}]{smallwood2012tracking}%
  \BibitemOpen
  \bibfield  {author} {\bibinfo {author} {\bibfnamefont {C.~L.}\ \bibnamefont
  {Smallwood}}, \bibinfo {author} {\bibfnamefont {J.~P.}\ \bibnamefont
  {Hinton}}, \bibinfo {author} {\bibfnamefont {C.}~\bibnamefont {Jozwiak}},
  \bibinfo {author} {\bibfnamefont {W.}~\bibnamefont {Zhang}}, \bibinfo
  {author} {\bibfnamefont {J.~D.}\ \bibnamefont {Koralek}}, \bibinfo {author}
  {\bibfnamefont {H.}~\bibnamefont {Eisaki}}, \bibinfo {author} {\bibfnamefont
  {D.-H.}\ \bibnamefont {Lee}}, \bibinfo {author} {\bibfnamefont
  {J.}~\bibnamefont {Orenstein}},\ and\ \bibinfo {author} {\bibfnamefont
  {A.}~\bibnamefont {Lanzara}},\ }\bibfield  {title} {\bibinfo {title}
  {Tracking cooper pairs in a cuprate superconductor by ultrafast
  angle-resolved photoemission},\ }\href@noop {} {\bibfield  {journal}
  {\bibinfo  {journal} {Science}\ }\textbf {\bibinfo {volume} {336}},\ \bibinfo
  {pages} {1137} (\bibinfo {year} {2012})}\BibitemShut {NoStop}%
\bibitem [{\citenamefont {Rameau}\ \emph {et~al.}(2014)\citenamefont {Rameau},
  \citenamefont {Freutel}, \citenamefont {Rettig}, \citenamefont {Avigo},
  \citenamefont {Ligges}, \citenamefont {Yoshida}, \citenamefont {Eisaki},
  \citenamefont {Schneeloch}, \citenamefont {Zhong}, \citenamefont {Xu} \emph
  {et~al.}}]{rameau2014photoinduced}%
  \BibitemOpen
  \bibfield  {author} {\bibinfo {author} {\bibfnamefont {J.}~\bibnamefont
  {Rameau}}, \bibinfo {author} {\bibfnamefont {S.}~\bibnamefont {Freutel}},
  \bibinfo {author} {\bibfnamefont {L.}~\bibnamefont {Rettig}}, \bibinfo
  {author} {\bibfnamefont {I.}~\bibnamefont {Avigo}}, \bibinfo {author}
  {\bibfnamefont {M.}~\bibnamefont {Ligges}}, \bibinfo {author} {\bibfnamefont
  {Y.}~\bibnamefont {Yoshida}}, \bibinfo {author} {\bibfnamefont
  {H.}~\bibnamefont {Eisaki}}, \bibinfo {author} {\bibfnamefont
  {J.}~\bibnamefont {Schneeloch}}, \bibinfo {author} {\bibfnamefont
  {R.}~\bibnamefont {Zhong}}, \bibinfo {author} {\bibfnamefont
  {Z.}~\bibnamefont {Xu}}, \emph {et~al.},\ }\bibfield  {title} {\bibinfo
  {title} {Photoinduced changes in the cuprate electronic structure revealed by
  femtosecond time-and angle-resolved photoemission},\ }\href@noop {}
  {\bibfield  {journal} {\bibinfo  {journal} {Phys. Rev. B}\ }\textbf {\bibinfo
  {volume} {89}},\ \bibinfo {pages} {115115} (\bibinfo {year}
  {2014})}\BibitemShut {NoStop}%
\bibitem [{\citenamefont {Graf}\ \emph {et~al.}(2011)\citenamefont {Graf},
  \citenamefont {Jozwiak}, \citenamefont {Smallwood}, \citenamefont {Eisaki},
  \citenamefont {Kaindl}, \citenamefont {Lee},\ and\ \citenamefont
  {Lanzara}}]{graf2011nodal}%
  \BibitemOpen
  \bibfield  {author} {\bibinfo {author} {\bibfnamefont {J.}~\bibnamefont
  {Graf}}, \bibinfo {author} {\bibfnamefont {C.}~\bibnamefont {Jozwiak}},
  \bibinfo {author} {\bibfnamefont {C.~L.}\ \bibnamefont {Smallwood}}, \bibinfo
  {author} {\bibfnamefont {H.}~\bibnamefont {Eisaki}}, \bibinfo {author}
  {\bibfnamefont {R.~A.}\ \bibnamefont {Kaindl}}, \bibinfo {author}
  {\bibfnamefont {D.-H.}\ \bibnamefont {Lee}},\ and\ \bibinfo {author}
  {\bibfnamefont {A.}~\bibnamefont {Lanzara}},\ }\bibfield  {title} {\bibinfo
  {title} {Nodal quasiparticle meltdown in ultrahigh-resolution pump--probe
  angle-resolved photoemission},\ }\href@noop {} {\bibfield  {journal}
  {\bibinfo  {journal} {Nat. Phys.}\ }\textbf {\bibinfo {volume} {7}},\
  \bibinfo {pages} {805} (\bibinfo {year} {2011})}\BibitemShut {NoStop}%
\bibitem [{\citenamefont {Perfetti}\ \emph {et~al.}(2007)\citenamefont
  {Perfetti}, \citenamefont {Loukakos}, \citenamefont {Lisowski}, \citenamefont
  {Bovensiepen}, \citenamefont {Eisaki},\ and\ \citenamefont
  {Wolf}}]{perfetti2007ultrafast}%
  \BibitemOpen
  \bibfield  {author} {\bibinfo {author} {\bibfnamefont {L.}~\bibnamefont
  {Perfetti}}, \bibinfo {author} {\bibfnamefont {P.}~\bibnamefont {Loukakos}},
  \bibinfo {author} {\bibfnamefont {M.}~\bibnamefont {Lisowski}}, \bibinfo
  {author} {\bibfnamefont {U.}~\bibnamefont {Bovensiepen}}, \bibinfo {author}
  {\bibfnamefont {H.}~\bibnamefont {Eisaki}},\ and\ \bibinfo {author}
  {\bibfnamefont {M.}~\bibnamefont {Wolf}},\ }\bibfield  {title} {\bibinfo
  {title} {Ultrafast electron relaxation in superconducting
  \ch{Bi2Sr2CaCu2O_{8+ $\delta$}} by time-resolved photoelectron
  spectroscopy},\ }\href@noop {} {\bibfield  {journal} {\bibinfo  {journal}
  {Phys. Rev. Lett.}\ }\textbf {\bibinfo {volume} {99}},\ \bibinfo {pages}
  {197001} (\bibinfo {year} {2007})}\BibitemShut {NoStop}%
\bibitem [{\citenamefont {Boschini}\ \emph {et~al.}(2018)\citenamefont
  {Boschini}, \citenamefont {da~Silva~Neto}, \citenamefont {Razzoli},
  \citenamefont {Zonno}, \citenamefont {Peli}, \citenamefont {Day},
  \citenamefont {Michiardi}, \citenamefont {Schneider}, \citenamefont
  {Zwartsenberg}, \citenamefont {Nigge} \emph {et~al.}}]{boschini2018collapse}%
  \BibitemOpen
  \bibfield  {author} {\bibinfo {author} {\bibfnamefont {F.}~\bibnamefont
  {Boschini}}, \bibinfo {author} {\bibfnamefont {E.}~\bibnamefont
  {da~Silva~Neto}}, \bibinfo {author} {\bibfnamefont {E.}~\bibnamefont
  {Razzoli}}, \bibinfo {author} {\bibfnamefont {M.}~\bibnamefont {Zonno}},
  \bibinfo {author} {\bibfnamefont {S.}~\bibnamefont {Peli}}, \bibinfo {author}
  {\bibfnamefont {R.}~\bibnamefont {Day}}, \bibinfo {author} {\bibfnamefont
  {M.}~\bibnamefont {Michiardi}}, \bibinfo {author} {\bibfnamefont
  {M.}~\bibnamefont {Schneider}}, \bibinfo {author} {\bibfnamefont
  {B.}~\bibnamefont {Zwartsenberg}}, \bibinfo {author} {\bibfnamefont
  {P.}~\bibnamefont {Nigge}}, \emph {et~al.},\ }\bibfield  {title} {\bibinfo
  {title} {Collapse of superconductivity in cuprates via ultrafast quenching of
  phase coherence},\ }\href@noop {} {\bibfield  {journal} {\bibinfo  {journal}
  {Nat. Mat.}\ }\textbf {\bibinfo {volume} {17}},\ \bibinfo {pages} {416}
  (\bibinfo {year} {2018})}\BibitemShut {NoStop}%
\bibitem [{\citenamefont {Zhang}\ \emph {et~al.}(2020)\citenamefont {Zhang},
  \citenamefont {Wang}, \citenamefont {Xiang}, \citenamefont {Yao},
  \citenamefont {Liu}, \citenamefont {Shi}, \citenamefont {Lin}, \citenamefont
  {Dong}, \citenamefont {Wu},\ and\ \citenamefont
  {Wang}}]{zhang2020photoinduced}%
  \BibitemOpen
  \bibfield  {author} {\bibinfo {author} {\bibfnamefont {S.}~\bibnamefont
  {Zhang}}, \bibinfo {author} {\bibfnamefont {Z.}~\bibnamefont {Wang}},
  \bibinfo {author} {\bibfnamefont {H.}~\bibnamefont {Xiang}}, \bibinfo
  {author} {\bibfnamefont {X.}~\bibnamefont {Yao}}, \bibinfo {author}
  {\bibfnamefont {Q.}~\bibnamefont {Liu}}, \bibinfo {author} {\bibfnamefont
  {L.}~\bibnamefont {Shi}}, \bibinfo {author} {\bibfnamefont {T.}~\bibnamefont
  {Lin}}, \bibinfo {author} {\bibfnamefont {T.}~\bibnamefont {Dong}}, \bibinfo
  {author} {\bibfnamefont {D.}~\bibnamefont {Wu}},\ and\ \bibinfo {author}
  {\bibfnamefont {N.}~\bibnamefont {Wang}},\ }\bibfield  {title} {\bibinfo
  {title} {Photoinduced nonequilibrium response in underdoped \ch{YBa 2 Cu 3
  O_{6+ x}} probed by time-resolved terahertz spectroscopy},\ }\href@noop {}
  {\bibfield  {journal} {\bibinfo  {journal} {Phys. Rev. X}\ }\textbf {\bibinfo
  {volume} {10}},\ \bibinfo {pages} {011056} (\bibinfo {year}
  {2020})}\BibitemShut {NoStop}%
\bibitem [{\citenamefont {Hellmann}\ \emph {et~al.}(2012)\citenamefont
  {Hellmann}, \citenamefont {Rohwer}, \citenamefont {Kall{\"a}ne},
  \citenamefont {Hanff}, \citenamefont {Sohrt}, \citenamefont {Stange},
  \citenamefont {Carr}, \citenamefont {Murnane}, \citenamefont {Kapteyn},
  \citenamefont {Kipp} \emph {et~al.}}]{hellmann2012time}%
  \BibitemOpen
  \bibfield  {author} {\bibinfo {author} {\bibfnamefont {S.}~\bibnamefont
  {Hellmann}}, \bibinfo {author} {\bibfnamefont {T.}~\bibnamefont {Rohwer}},
  \bibinfo {author} {\bibfnamefont {M.}~\bibnamefont {Kall{\"a}ne}}, \bibinfo
  {author} {\bibfnamefont {K.}~\bibnamefont {Hanff}}, \bibinfo {author}
  {\bibfnamefont {C.}~\bibnamefont {Sohrt}}, \bibinfo {author} {\bibfnamefont
  {A.}~\bibnamefont {Stange}}, \bibinfo {author} {\bibfnamefont
  {A.}~\bibnamefont {Carr}}, \bibinfo {author} {\bibfnamefont {M.}~\bibnamefont
  {Murnane}}, \bibinfo {author} {\bibfnamefont {H.}~\bibnamefont {Kapteyn}},
  \bibinfo {author} {\bibfnamefont {L.}~\bibnamefont {Kipp}}, \emph {et~al.},\
  }\bibfield  {title} {\bibinfo {title} {Time-domain classification of
  charge-density-wave insulators},\ }\href@noop {} {\bibfield  {journal}
  {\bibinfo  {journal} {Nat. Comm.}\ }\textbf {\bibinfo {volume} {3}},\
  \bibinfo {pages} {1069} (\bibinfo {year} {2012})}\BibitemShut {NoStop}%
\bibitem [{\citenamefont {Sobota}\ \emph {et~al.}(2021)\citenamefont {Sobota},
  \citenamefont {He},\ and\ \citenamefont {Shen}}]{sobota2021angle}%
  \BibitemOpen
  \bibfield  {author} {\bibinfo {author} {\bibfnamefont {J.~A.}\ \bibnamefont
  {Sobota}}, \bibinfo {author} {\bibfnamefont {Y.}~\bibnamefont {He}},\ and\
  \bibinfo {author} {\bibfnamefont {Z.-X.}\ \bibnamefont {Shen}},\ }\bibfield
  {title} {\bibinfo {title} {Angle-resolved photoemission studies of quantum
  materials},\ }\href@noop {} {\bibfield  {journal} {\bibinfo  {journal} {Rev.
  Mod. Phys.}\ }\textbf {\bibinfo {volume} {93}},\ \bibinfo {pages} {025006}
  (\bibinfo {year} {2021})}\BibitemShut {NoStop}%
\bibitem [{\citenamefont {Yang}\ \emph {et~al.}(2015)\citenamefont {Yang},
  \citenamefont {Sobota}, \citenamefont {Leuenberger}, \citenamefont {He},
  \citenamefont {Hashimoto}, \citenamefont {Lu}, \citenamefont {Eisaki},
  \citenamefont {Kirchmann},\ and\ \citenamefont
  {Shen}}]{yang2015inequivalence}%
  \BibitemOpen
  \bibfield  {author} {\bibinfo {author} {\bibfnamefont {S.-L.}\ \bibnamefont
  {Yang}}, \bibinfo {author} {\bibfnamefont {J.~A.}\ \bibnamefont {Sobota}},
  \bibinfo {author} {\bibfnamefont {D.}~\bibnamefont {Leuenberger}}, \bibinfo
  {author} {\bibfnamefont {Y.}~\bibnamefont {He}}, \bibinfo {author}
  {\bibfnamefont {M.}~\bibnamefont {Hashimoto}}, \bibinfo {author}
  {\bibfnamefont {D.}~\bibnamefont {Lu}}, \bibinfo {author} {\bibfnamefont
  {H.}~\bibnamefont {Eisaki}}, \bibinfo {author} {\bibfnamefont {P.~S.}\
  \bibnamefont {Kirchmann}},\ and\ \bibinfo {author} {\bibfnamefont {Z.-X.}\
  \bibnamefont {Shen}},\ }\bibfield  {title} {\bibinfo {title} {Inequivalence
  of single-particle and population lifetimes in a cuprate superconductor},\
  }\href@noop {} {\bibfield  {journal} {\bibinfo  {journal} {Phys. Rev. Lett.}\
  }\textbf {\bibinfo {volume} {114}},\ \bibinfo {pages} {247001} (\bibinfo
  {year} {2015})}\BibitemShut {NoStop}%
\bibitem [{\citenamefont {Cilento}\ \emph {et~al.}(2018)\citenamefont
  {Cilento}, \citenamefont {Manzoni}, \citenamefont {Sterzi}, \citenamefont
  {Peli}, \citenamefont {Ronchi}, \citenamefont {Crepaldi}, \citenamefont
  {Boschini}, \citenamefont {Cacho}, \citenamefont {Chapman}, \citenamefont
  {Springate} \emph {et~al.}}]{cilento2018dynamics}%
  \BibitemOpen
  \bibfield  {author} {\bibinfo {author} {\bibfnamefont {F.}~\bibnamefont
  {Cilento}}, \bibinfo {author} {\bibfnamefont {G.}~\bibnamefont {Manzoni}},
  \bibinfo {author} {\bibfnamefont {A.}~\bibnamefont {Sterzi}}, \bibinfo
  {author} {\bibfnamefont {S.}~\bibnamefont {Peli}}, \bibinfo {author}
  {\bibfnamefont {A.}~\bibnamefont {Ronchi}}, \bibinfo {author} {\bibfnamefont
  {A.}~\bibnamefont {Crepaldi}}, \bibinfo {author} {\bibfnamefont
  {F.}~\bibnamefont {Boschini}}, \bibinfo {author} {\bibfnamefont
  {C.}~\bibnamefont {Cacho}}, \bibinfo {author} {\bibfnamefont
  {R.}~\bibnamefont {Chapman}}, \bibinfo {author} {\bibfnamefont
  {E.}~\bibnamefont {Springate}}, \emph {et~al.},\ }\bibfield  {title}
  {\bibinfo {title} {Dynamics of correlation-frozen antinodal quasiparticles in
  superconducting cuprates},\ }\href@noop {} {\bibfield  {journal} {\bibinfo
  {journal} {Sci. Adv.}\ }\textbf {\bibinfo {volume} {4}},\ \bibinfo {pages}
  {eaar1998} (\bibinfo {year} {2018})}\BibitemShut {NoStop}%
\bibitem [{\citenamefont {Sansone}\ \emph {et~al.}(2011)\citenamefont
  {Sansone}, \citenamefont {Poletto},\ and\ \citenamefont
  {Nisoli}}]{sansone2011high}%
  \BibitemOpen
  \bibfield  {author} {\bibinfo {author} {\bibfnamefont {G.}~\bibnamefont
  {Sansone}}, \bibinfo {author} {\bibfnamefont {L.}~\bibnamefont {Poletto}},\
  and\ \bibinfo {author} {\bibfnamefont {M.}~\bibnamefont {Nisoli}},\
  }\bibfield  {title} {\bibinfo {title} {High-energy attosecond light
  sources},\ }\href@noop {} {\bibfield  {journal} {\bibinfo  {journal} {Nat.
  Photonics}\ }\textbf {\bibinfo {volume} {5}},\ \bibinfo {pages} {655}
  (\bibinfo {year} {2011})}\BibitemShut {NoStop}%
\bibitem [{\citenamefont {Li}\ \emph {et~al.}(2020)\citenamefont {Li},
  \citenamefont {Lu}, \citenamefont {Chew}, \citenamefont {Han}, \citenamefont
  {Li}, \citenamefont {Wu}, \citenamefont {Wang}, \citenamefont {Ghimire},\
  and\ \citenamefont {Chang}}]{li2020attosecond}%
  \BibitemOpen
  \bibfield  {author} {\bibinfo {author} {\bibfnamefont {J.}~\bibnamefont
  {Li}}, \bibinfo {author} {\bibfnamefont {J.}~\bibnamefont {Lu}}, \bibinfo
  {author} {\bibfnamefont {A.}~\bibnamefont {Chew}}, \bibinfo {author}
  {\bibfnamefont {S.}~\bibnamefont {Han}}, \bibinfo {author} {\bibfnamefont
  {J.}~\bibnamefont {Li}}, \bibinfo {author} {\bibfnamefont {Y.}~\bibnamefont
  {Wu}}, \bibinfo {author} {\bibfnamefont {H.}~\bibnamefont {Wang}}, \bibinfo
  {author} {\bibfnamefont {S.}~\bibnamefont {Ghimire}},\ and\ \bibinfo {author}
  {\bibfnamefont {Z.}~\bibnamefont {Chang}},\ }\bibfield  {title} {\bibinfo
  {title} {Attosecond science based on high harmonic generation from gases and
  solids},\ }\href@noop {} {\bibfield  {journal} {\bibinfo  {journal} {Nat.
  Comm.}\ }\textbf {\bibinfo {volume} {11}},\ \bibinfo {pages} {2748} (\bibinfo
  {year} {2020})}\BibitemShut {NoStop}%
\bibitem [{\citenamefont {Giannetti}\ \emph {et~al.}(2009)\citenamefont
  {Giannetti}, \citenamefont {Coslovich}, \citenamefont {Cilento},
  \citenamefont {Ferrini}, \citenamefont {Eisaki}, \citenamefont {Kaneko},
  \citenamefont {Greven},\ and\ \citenamefont
  {Parmigiani}}]{giannetti2009discontinuity}%
  \BibitemOpen
  \bibfield  {author} {\bibinfo {author} {\bibfnamefont {C.}~\bibnamefont
  {Giannetti}}, \bibinfo {author} {\bibfnamefont {G.}~\bibnamefont
  {Coslovich}}, \bibinfo {author} {\bibfnamefont {F.}~\bibnamefont {Cilento}},
  \bibinfo {author} {\bibfnamefont {G.}~\bibnamefont {Ferrini}}, \bibinfo
  {author} {\bibfnamefont {H.}~\bibnamefont {Eisaki}}, \bibinfo {author}
  {\bibfnamefont {N.}~\bibnamefont {Kaneko}}, \bibinfo {author} {\bibfnamefont
  {M.}~\bibnamefont {Greven}},\ and\ \bibinfo {author} {\bibfnamefont
  {F.}~\bibnamefont {Parmigiani}},\ }\bibfield  {title} {\bibinfo {title}
  {{Discontinuity of the ultrafast electronic response of underdoped
  superconducting
  ${\text{Bi}}_{2}{\text{Sr}}_{2}{\text{CaCu}}_{2}{\text{O}}_{8+\ensuremath{\delta}}$
  strongly excited by ultrashort light pulses}},\ }\href
  {https://doi.org/10.1103/PhysRevB.79.224502} {\bibfield  {journal} {\bibinfo
  {journal} {Phys. Rev. B}\ }\textbf {\bibinfo {volume} {79}},\ \bibinfo
  {pages} {224502} (\bibinfo {year} {2009})}\BibitemShut {NoStop}%
\bibitem [{sup()}]{supplemental}%
  \BibitemOpen
  \href@noop {} {}\bibinfo {note} {See Supplemental Material at [URL] for an
  experimental setup schematic, details of the penetration-depth-dependent
  transient reflectivity model, fits to all EUV pump–optical probe transient
  reflectivity data, extracted fit parameters, and fits and fluence dependence
  of optical pump data.}\BibitemShut {Stop}%
\bibitem [{\citenamefont {Allaria}\ \emph {et~al.}(2015)\citenamefont
  {Allaria}, \citenamefont {Badano}, \citenamefont {Bassanese}, \citenamefont
  {Capotondi}, \citenamefont {Castronovo}, \citenamefont {Cinquegrana},
  \citenamefont {Danailov}, \citenamefont {D'auria}, \citenamefont
  {Demidovich}, \citenamefont {De~Monte} \emph {et~al.}}]{allaria2015fermi}%
  \BibitemOpen
  \bibfield  {author} {\bibinfo {author} {\bibfnamefont {E.}~\bibnamefont
  {Allaria}}, \bibinfo {author} {\bibfnamefont {L.}~\bibnamefont {Badano}},
  \bibinfo {author} {\bibfnamefont {S.}~\bibnamefont {Bassanese}}, \bibinfo
  {author} {\bibfnamefont {F.}~\bibnamefont {Capotondi}}, \bibinfo {author}
  {\bibfnamefont {D.}~\bibnamefont {Castronovo}}, \bibinfo {author}
  {\bibfnamefont {P.}~\bibnamefont {Cinquegrana}}, \bibinfo {author}
  {\bibfnamefont {M.}~\bibnamefont {Danailov}}, \bibinfo {author}
  {\bibfnamefont {G.}~\bibnamefont {D'auria}}, \bibinfo {author} {\bibfnamefont
  {A.}~\bibnamefont {Demidovich}}, \bibinfo {author} {\bibfnamefont
  {R.}~\bibnamefont {De~Monte}}, \emph {et~al.},\ }\bibfield  {title} {\bibinfo
  {title} {{The FERMI free-electron lasers}},\ }\href@noop {} {\bibfield
  {journal} {\bibinfo  {journal} {Synchrotron Radiation}\ }\textbf {\bibinfo
  {volume} {22}},\ \bibinfo {pages} {485} (\bibinfo {year} {2015})}\BibitemShut
  {NoStop}%
\bibitem [{\citenamefont {Finetti}\ \emph {et~al.}(2017)\citenamefont
  {Finetti}, \citenamefont {H{\"o}ppner}, \citenamefont {Allaria},
  \citenamefont {Callegari}, \citenamefont {Capotondi}, \citenamefont
  {Cinquegrana}, \citenamefont {Coreno}, \citenamefont {Cucini}, \citenamefont
  {Danailov}, \citenamefont {Demidovich} \emph {et~al.}}]{finetti2017pulse}%
  \BibitemOpen
  \bibfield  {author} {\bibinfo {author} {\bibfnamefont {P.}~\bibnamefont
  {Finetti}}, \bibinfo {author} {\bibfnamefont {H.}~\bibnamefont
  {H{\"o}ppner}}, \bibinfo {author} {\bibfnamefont {E.}~\bibnamefont
  {Allaria}}, \bibinfo {author} {\bibfnamefont {C.}~\bibnamefont {Callegari}},
  \bibinfo {author} {\bibfnamefont {F.}~\bibnamefont {Capotondi}}, \bibinfo
  {author} {\bibfnamefont {P.}~\bibnamefont {Cinquegrana}}, \bibinfo {author}
  {\bibfnamefont {M.}~\bibnamefont {Coreno}}, \bibinfo {author} {\bibfnamefont
  {R.}~\bibnamefont {Cucini}}, \bibinfo {author} {\bibfnamefont {M.~B.}\
  \bibnamefont {Danailov}}, \bibinfo {author} {\bibfnamefont {A.}~\bibnamefont
  {Demidovich}}, \emph {et~al.},\ }\bibfield  {title} {\bibinfo {title} {Pulse
  duration of seeded free-electron lasers},\ }\href@noop {} {\bibfield
  {journal} {\bibinfo  {journal} {Phys. Rev. X}\ }\textbf {\bibinfo {volume}
  {7}},\ \bibinfo {pages} {021043} (\bibinfo {year} {2017})}\BibitemShut
  {NoStop}%
\bibitem [{\citenamefont {Junod}\ \emph {et~al.}(1994)\citenamefont {Junod},
  \citenamefont {Wang}, \citenamefont {Tsukamoto}, \citenamefont {Triscone},
  \citenamefont {Revaz}, \citenamefont {Walker},\ and\ \citenamefont
  {Muller}}]{junod1994specific}%
  \BibitemOpen
  \bibfield  {author} {\bibinfo {author} {\bibfnamefont {A.}~\bibnamefont
  {Junod}}, \bibinfo {author} {\bibfnamefont {K.-Q.}\ \bibnamefont {Wang}},
  \bibinfo {author} {\bibfnamefont {T.}~\bibnamefont {Tsukamoto}}, \bibinfo
  {author} {\bibfnamefont {G.}~\bibnamefont {Triscone}}, \bibinfo {author}
  {\bibfnamefont {B.}~\bibnamefont {Revaz}}, \bibinfo {author} {\bibfnamefont
  {E.}~\bibnamefont {Walker}},\ and\ \bibinfo {author} {\bibfnamefont
  {J.}~\bibnamefont {Muller}},\ }\bibfield  {title} {\bibinfo {title} {Specific
  heat up to 14 tesla and magnetization of a \ch{Bi2Sr2CaCu2O8} single crystal
  thermodynamics of a 2d superconductor},\ }\href@noop {} {\bibfield  {journal}
  {\bibinfo  {journal} {Physica C: Superconductivity}\ }\textbf {\bibinfo
  {volume} {229}},\ \bibinfo {pages} {209} (\bibinfo {year}
  {1994})}\BibitemShut {NoStop}%
\bibitem [{\citenamefont {Dal~Conte}\ \emph {et~al.}(2012)\citenamefont
  {Dal~Conte}, \citenamefont {Giannetti}, \citenamefont {Coslovich},
  \citenamefont {Cilento}, \citenamefont {Bossini}, \citenamefont {Abebaw},
  \citenamefont {Banfi}, \citenamefont {Ferrini}, \citenamefont {Eisaki},
  \citenamefont {Greven} \emph {et~al.}}]{dal2012disentangling}%
  \BibitemOpen
  \bibfield  {author} {\bibinfo {author} {\bibfnamefont {S.}~\bibnamefont
  {Dal~Conte}}, \bibinfo {author} {\bibfnamefont {C.}~\bibnamefont
  {Giannetti}}, \bibinfo {author} {\bibfnamefont {G.}~\bibnamefont
  {Coslovich}}, \bibinfo {author} {\bibfnamefont {F.}~\bibnamefont {Cilento}},
  \bibinfo {author} {\bibfnamefont {D.}~\bibnamefont {Bossini}}, \bibinfo
  {author} {\bibfnamefont {T.}~\bibnamefont {Abebaw}}, \bibinfo {author}
  {\bibfnamefont {F.}~\bibnamefont {Banfi}}, \bibinfo {author} {\bibfnamefont
  {G.}~\bibnamefont {Ferrini}}, \bibinfo {author} {\bibfnamefont
  {H.}~\bibnamefont {Eisaki}}, \bibinfo {author} {\bibfnamefont
  {M.}~\bibnamefont {Greven}}, \emph {et~al.},\ }\bibfield  {title} {\bibinfo
  {title} {{Disentangling the electronic and phononic glue in a high-$T_c$
  superconductor}},\ }\href@noop {} {\bibfield  {journal} {\bibinfo  {journal}
  {Science}\ }\textbf {\bibinfo {volume} {335}},\ \bibinfo {pages} {1600}
  (\bibinfo {year} {2012})}\BibitemShut {NoStop}%
\bibitem [{\citenamefont {Giannetti}\ \emph {et~al.}(2011)\citenamefont
  {Giannetti}, \citenamefont {Cilento}, \citenamefont {Conte}, \citenamefont
  {Coslovich}, \citenamefont {Ferrini}, \citenamefont {Molegraaf},
  \citenamefont {Raichle}, \citenamefont {Liang}, \citenamefont {Eisaki},
  \citenamefont {Greven} \emph {et~al.}}]{giannetti2011revealing}%
  \BibitemOpen
  \bibfield  {author} {\bibinfo {author} {\bibfnamefont {C.}~\bibnamefont
  {Giannetti}}, \bibinfo {author} {\bibfnamefont {F.}~\bibnamefont {Cilento}},
  \bibinfo {author} {\bibfnamefont {S.~D.}\ \bibnamefont {Conte}}, \bibinfo
  {author} {\bibfnamefont {G.}~\bibnamefont {Coslovich}}, \bibinfo {author}
  {\bibfnamefont {G.}~\bibnamefont {Ferrini}}, \bibinfo {author} {\bibfnamefont
  {H.}~\bibnamefont {Molegraaf}}, \bibinfo {author} {\bibfnamefont
  {M.}~\bibnamefont {Raichle}}, \bibinfo {author} {\bibfnamefont
  {R.}~\bibnamefont {Liang}}, \bibinfo {author} {\bibfnamefont
  {H.}~\bibnamefont {Eisaki}}, \bibinfo {author} {\bibfnamefont
  {M.}~\bibnamefont {Greven}}, \emph {et~al.},\ }\bibfield  {title} {\bibinfo
  {title} {Revealing the high-energy electronic excitations underlying the
  onset of high-temperature superconductivity in cuprates},\ }\href@noop {}
  {\bibfield  {journal} {\bibinfo  {journal} {Nat. Comm.}\ }\textbf {\bibinfo
  {volume} {2}},\ \bibinfo {pages} {353} (\bibinfo {year} {2011})}\BibitemShut
  {NoStop}%
\bibitem [{CXR()}]{CXRO}%
  \BibitemOpen
  \href@noop {} {\bibinfo {title} {{The Center for X-Ray Optics (CXRO)}}},\
  \bibinfo {howpublished}
  {\url{https://henke.lbl.gov/optical_constants/}}\BibitemShut {NoStop}%
\bibitem [{\citenamefont {Henke}\ \emph {et~al.}(1993)\citenamefont {Henke},
  \citenamefont {Gullikson},\ and\ \citenamefont {Davis}}]{Henke1993}%
  \BibitemOpen
  \bibfield  {author} {\bibinfo {author} {\bibfnamefont {B.}~\bibnamefont
  {Henke}}, \bibinfo {author} {\bibfnamefont {E.}~\bibnamefont {Gullikson}},\
  and\ \bibinfo {author} {\bibfnamefont {J.}~\bibnamefont {Davis}},\ }\bibfield
   {title} {\bibinfo {title} {{X-Ray Interactions: Photoabsorption, Scattering,
  Transmission, and Reflection at E = 50-30,000 eV, Z = 1-92}},\ }\href
  {https://doi.org/https://doi.org/10.1006/adnd.1993.1013} {\bibfield
  {journal} {\bibinfo  {journal} {Atomic Data and Nuclear Data Tables}\
  }\textbf {\bibinfo {volume} {54}},\ \bibinfo {pages} {181} (\bibinfo {year}
  {1993})}\BibitemShut {NoStop}%
\bibitem [{\citenamefont {Allen}(1987)}]{allen1987theory}%
  \BibitemOpen
  \bibfield  {author} {\bibinfo {author} {\bibfnamefont {P.~B.}\ \bibnamefont
  {Allen}},\ }\bibfield  {title} {\bibinfo {title} {Theory of thermal
  relaxation of electrons in metals},\ }\href@noop {} {\bibfield  {journal}
  {\bibinfo  {journal} {Phys. Rev. Lett.}\ }\textbf {\bibinfo {volume} {59}},\
  \bibinfo {pages} {1460} (\bibinfo {year} {1987})}\BibitemShut {NoStop}%
\bibitem [{\citenamefont {Thomsen}\ \emph {et~al.}(1986)\citenamefont
  {Thomsen}, \citenamefont {Grahn}, \citenamefont {Maris},\ and\ \citenamefont
  {Tauc}}]{thomsen1986surface}%
  \BibitemOpen
  \bibfield  {author} {\bibinfo {author} {\bibfnamefont {C.}~\bibnamefont
  {Thomsen}}, \bibinfo {author} {\bibfnamefont {H.~T.}\ \bibnamefont {Grahn}},
  \bibinfo {author} {\bibfnamefont {H.~J.}\ \bibnamefont {Maris}},\ and\
  \bibinfo {author} {\bibfnamefont {J.}~\bibnamefont {Tauc}},\ }\bibfield
  {title} {\bibinfo {title} {Surface generation and detection of phonons by
  picosecond light pulses},\ }\href@noop {} {\bibfield  {journal} {\bibinfo
  {journal} {Phys. Rev. B}\ }\textbf {\bibinfo {volume} {34}},\ \bibinfo
  {pages} {4129} (\bibinfo {year} {1986})}\BibitemShut {NoStop}%
\bibitem [{\citenamefont {Saunders}\ \emph {et~al.}(1994)\citenamefont
  {Saunders}, \citenamefont {Fanggao}, \citenamefont {Jiaqiang}, \citenamefont
  {Wang}, \citenamefont {Cankurtaran}, \citenamefont {Lambson}, \citenamefont
  {Ford},\ and\ \citenamefont {Almond}}]{saunders1994anisotropy}%
  \BibitemOpen
  \bibfield  {author} {\bibinfo {author} {\bibfnamefont {G.}~\bibnamefont
  {Saunders}}, \bibinfo {author} {\bibfnamefont {C.}~\bibnamefont {Fanggao}},
  \bibinfo {author} {\bibfnamefont {L.}~\bibnamefont {Jiaqiang}}, \bibinfo
  {author} {\bibfnamefont {Q.}~\bibnamefont {Wang}}, \bibinfo {author}
  {\bibfnamefont {M.}~\bibnamefont {Cankurtaran}}, \bibinfo {author}
  {\bibfnamefont {E.}~\bibnamefont {Lambson}}, \bibinfo {author} {\bibfnamefont
  {P.}~\bibnamefont {Ford}},\ and\ \bibinfo {author} {\bibfnamefont
  {D.}~\bibnamefont {Almond}},\ }\bibfield  {title} {\bibinfo {title}
  {Anisotropy of the elastic and nonlinear acoustic properties of dense
  textured \ch{Bi2Sr2CaCu2O_{8+ y}}},\ }\href@noop {} {\bibfield  {journal}
  {\bibinfo  {journal} {Phys. Rev. B}\ }\textbf {\bibinfo {volume} {49}},\
  \bibinfo {pages} {9862} (\bibinfo {year} {1994})}\BibitemShut {NoStop}%
\bibitem [{\citenamefont {Wang}\ \emph {et~al.}(1989)\citenamefont {Wang},
  \citenamefont {Wu}, \citenamefont {Zhu}, \citenamefont {Shen}, \citenamefont
  {Zhang}, \citenamefont {Yan},\ and\ \citenamefont {Zhao}}]{wang1989elastic}%
  \BibitemOpen
  \bibfield  {author} {\bibinfo {author} {\bibfnamefont {Y.}~\bibnamefont
  {Wang}}, \bibinfo {author} {\bibfnamefont {J.}~\bibnamefont {Wu}}, \bibinfo
  {author} {\bibfnamefont {J.}~\bibnamefont {Zhu}}, \bibinfo {author}
  {\bibfnamefont {H.}~\bibnamefont {Shen}}, \bibinfo {author} {\bibfnamefont
  {J.}~\bibnamefont {Zhang}}, \bibinfo {author} {\bibfnamefont
  {Y.}~\bibnamefont {Yan}},\ and\ \bibinfo {author} {\bibfnamefont
  {Z.}~\bibnamefont {Zhao}},\ }\bibfield  {title} {\bibinfo {title} {Elastic
  anisotropy and lattice instability in \ch{Bi2Sr2Ca1Cu2O8} single crystal},\
  }\href@noop {} {\bibfield  {journal} {\bibinfo  {journal} {Phys. Lett. A}\
  }\textbf {\bibinfo {volume} {142}},\ \bibinfo {pages} {289} (\bibinfo {year}
  {1989})}\BibitemShut {NoStop}%
\bibitem [{\citenamefont {McNiven}\ \emph {et~al.}(2022)\citenamefont
  {McNiven}, \citenamefont {LeBlanc},\ and\ \citenamefont
  {Andrews}}]{mcniven2022long}%
  \BibitemOpen
  \bibfield  {author} {\bibinfo {author} {\bibfnamefont {B.~D.}\ \bibnamefont
  {McNiven}}, \bibinfo {author} {\bibfnamefont {J.~P.}\ \bibnamefont
  {LeBlanc}},\ and\ \bibinfo {author} {\bibfnamefont {G.~T.}\ \bibnamefont
  {Andrews}},\ }\bibfield  {title} {\bibinfo {title} {{Long-wavelength phonon
  dynamics in incommensurate \ch{Bi2Sr2CaCu2O_{8+ $\delta$}} crystals by
  Brillouin light scattering spectroscopy}},\ }\href@noop {} {\bibfield
  {journal} {\bibinfo  {journal} {Phys. Rev. B}\ }\textbf {\bibinfo {volume}
  {106}},\ \bibinfo {pages} {054113} (\bibinfo {year} {2022})}\BibitemShut
  {NoStop}%
\bibitem [{\citenamefont {Chen}\ \emph {et~al.}(2019)\citenamefont {Chen},
  \citenamefont {Wang}, \citenamefont {Jia}, \citenamefont {Moritz},
  \citenamefont {Shvaika}, \citenamefont {Freericks},\ and\ \citenamefont
  {Devereaux}}]{Chen2019trRIXS}%
  \BibitemOpen
  \bibfield  {author} {\bibinfo {author} {\bibfnamefont {Y.}~\bibnamefont
  {Chen}}, \bibinfo {author} {\bibfnamefont {Y.}~\bibnamefont {Wang}}, \bibinfo
  {author} {\bibfnamefont {C.}~\bibnamefont {Jia}}, \bibinfo {author}
  {\bibfnamefont {B.}~\bibnamefont {Moritz}}, \bibinfo {author} {\bibfnamefont
  {A.~M.}\ \bibnamefont {Shvaika}}, \bibinfo {author} {\bibfnamefont {J.~K.}\
  \bibnamefont {Freericks}},\ and\ \bibinfo {author} {\bibfnamefont {T.~P.}\
  \bibnamefont {Devereaux}},\ }\bibfield  {title} {\bibinfo {title} {Theory for
  time-resolved resonant inelastic x-ray scattering},\ }\href
  {https://doi.org/10.1103/PhysRevB.99.104306} {\ \textbf {\bibinfo {volume}
  {99}},\ \bibinfo {pages} {104306} (\bibinfo {year} {2019})}\BibitemShut
  {NoStop}%
\bibitem [{\citenamefont {Mukamel}\ \emph {et~al.}(2013)\citenamefont
  {Mukamel}, \citenamefont {Healion}, \citenamefont {Zhang},\ and\
  \citenamefont {Biggs}}]{Mukamel2013}%
  \BibitemOpen
  \bibfield  {author} {\bibinfo {author} {\bibfnamefont {S.}~\bibnamefont
  {Mukamel}}, \bibinfo {author} {\bibfnamefont {D.}~\bibnamefont {Healion}},
  \bibinfo {author} {\bibfnamefont {Y.}~\bibnamefont {Zhang}},\ and\ \bibinfo
  {author} {\bibfnamefont {J.~D.}\ \bibnamefont {Biggs}},\ }\bibfield  {title}
  {\bibinfo {title} {{Multidimensional attosecond resonant X-ray spectroscopy
  of molecules: lessons from the optical regime}},\ }\href
  {https://doi.org/10.1146/annurev-physchem-040412-110021} {\bibfield
  {journal} {\bibinfo  {journal} {Annual Review of Physical Chemistry}\
  }\textbf {\bibinfo {volume} {64}},\ \bibinfo {pages} {101} (\bibinfo {year}
  {2013})}\BibitemShut {NoStop}%
\bibitem [{\citenamefont {Madan}\ \emph {et~al.}(2015)\citenamefont {Madan},
  \citenamefont {Kurosawa}, \citenamefont {Toda}, \citenamefont {Oda},
  \citenamefont {Mertelj},\ and\ \citenamefont {Mihailovic}}]{Madan2014}%
  \BibitemOpen
  \bibfield  {author} {\bibinfo {author} {\bibfnamefont {I.}~\bibnamefont
  {Madan}}, \bibinfo {author} {\bibfnamefont {T.}~\bibnamefont {Kurosawa}},
  \bibinfo {author} {\bibfnamefont {Y.}~\bibnamefont {Toda}}, \bibinfo {author}
  {\bibfnamefont {M.}~\bibnamefont {Oda}}, \bibinfo {author} {\bibfnamefont
  {T.}~\bibnamefont {Mertelj}},\ and\ \bibinfo {author} {\bibfnamefont
  {D.}~\bibnamefont {Mihailovic}},\ }\bibfield  {title} {\bibinfo {title}
  {Evidence for carrier localization in the pseudogap state of cuprate
  superconductors from coherent quench experiments},\ }\href@noop {} {\bibfield
   {journal} {\bibinfo  {journal} {Nat. Comm.}\ }\textbf {\bibinfo {volume}
  {6}},\ \bibinfo {pages} {6958} (\bibinfo {year} {2015})}\BibitemShut
  {NoStop}%
\bibitem [{\citenamefont {Sidiropoulos}\ \emph {et~al.}(2021)\citenamefont
  {Sidiropoulos}, \citenamefont {Di~Palo}, \citenamefont {Rivas}, \citenamefont
  {Severino}, \citenamefont {Reduzzi}, \citenamefont {Nandy}, \citenamefont
  {Bauerhenne}, \citenamefont {Krylow}, \citenamefont {Vasileiadis},
  \citenamefont {Danz}, \citenamefont {Elliott}, \citenamefont {Sharma},
  \citenamefont {Dewhurst}, \citenamefont {Ropers}, \citenamefont {Joly},
  \citenamefont {Garcia}, \citenamefont {Wolf}, \citenamefont {Ernstorfer},\
  and\ \citenamefont {Biegert}}]{Sidiropoulos2021}%
  \BibitemOpen
  \bibfield  {author} {\bibinfo {author} {\bibfnamefont {T.~P.~H.}\
  \bibnamefont {Sidiropoulos}}, \bibinfo {author} {\bibfnamefont
  {N.}~\bibnamefont {Di~Palo}}, \bibinfo {author} {\bibfnamefont {D.~E.}\
  \bibnamefont {Rivas}}, \bibinfo {author} {\bibfnamefont {S.}~\bibnamefont
  {Severino}}, \bibinfo {author} {\bibfnamefont {M.}~\bibnamefont {Reduzzi}},
  \bibinfo {author} {\bibfnamefont {B.}~\bibnamefont {Nandy}}, \bibinfo
  {author} {\bibfnamefont {B.}~\bibnamefont {Bauerhenne}}, \bibinfo {author}
  {\bibfnamefont {S.}~\bibnamefont {Krylow}}, \bibinfo {author} {\bibfnamefont
  {T.}~\bibnamefont {Vasileiadis}}, \bibinfo {author} {\bibfnamefont
  {T.}~\bibnamefont {Danz}}, \bibinfo {author} {\bibfnamefont {P.}~\bibnamefont
  {Elliott}}, \bibinfo {author} {\bibfnamefont {S.}~\bibnamefont {Sharma}},
  \bibinfo {author} {\bibfnamefont {K.}~\bibnamefont {Dewhurst}}, \bibinfo
  {author} {\bibfnamefont {C.}~\bibnamefont {Ropers}}, \bibinfo {author}
  {\bibfnamefont {Y.}~\bibnamefont {Joly}}, \bibinfo {author} {\bibfnamefont
  {M.~E.}\ \bibnamefont {Garcia}}, \bibinfo {author} {\bibfnamefont
  {M.}~\bibnamefont {Wolf}}, \bibinfo {author} {\bibfnamefont {R.}~\bibnamefont
  {Ernstorfer}},\ and\ \bibinfo {author} {\bibfnamefont {J.}~\bibnamefont
  {Biegert}},\ }\bibfield  {title} {\bibinfo {title} {Probing the energy
  conversion pathways between light, carriers, and lattice in real time with
  attosecond core-level spectroscopy},\ }\href
  {https://doi.org/10.1103/PhysRevX.11.041060} {\bibfield  {journal} {\bibinfo
  {journal} {Phys. Rev. X}\ }\textbf {\bibinfo {volume} {11}},\ \bibinfo
  {pages} {041060} (\bibinfo {year} {2021})}\BibitemShut {NoStop}%
\bibitem [{\citenamefont {Seddon}\ \emph {et~al.}(2017)\citenamefont {Seddon},
  \citenamefont {Clarke}, \citenamefont {Dunning}, \citenamefont
  {Masciovecchio}, \citenamefont {Milne}, \citenamefont {Parmigiani},
  \citenamefont {Rugg}, \citenamefont {Spence}, \citenamefont {Thompson},
  \citenamefont {Ueda}, \citenamefont {Vinko}, \citenamefont {Wark},\ and\
  \citenamefont {Wurth}}]{Seddon2017}%
  \BibitemOpen
  \bibfield  {author} {\bibinfo {author} {\bibfnamefont {E.~A.}\ \bibnamefont
  {Seddon}}, \bibinfo {author} {\bibfnamefont {J.~A.}\ \bibnamefont {Clarke}},
  \bibinfo {author} {\bibfnamefont {D.~J.}\ \bibnamefont {Dunning}}, \bibinfo
  {author} {\bibfnamefont {C.}~\bibnamefont {Masciovecchio}}, \bibinfo {author}
  {\bibfnamefont {C.~J.}\ \bibnamefont {Milne}}, \bibinfo {author}
  {\bibfnamefont {F.}~\bibnamefont {Parmigiani}}, \bibinfo {author}
  {\bibfnamefont {D.}~\bibnamefont {Rugg}}, \bibinfo {author} {\bibfnamefont
  {J.~C.~H.}\ \bibnamefont {Spence}}, \bibinfo {author} {\bibfnamefont {N.~R.}\
  \bibnamefont {Thompson}}, \bibinfo {author} {\bibfnamefont {K.}~\bibnamefont
  {Ueda}}, \bibinfo {author} {\bibfnamefont {S.~M.}\ \bibnamefont {Vinko}},
  \bibinfo {author} {\bibfnamefont {J.~S.}\ \bibnamefont {Wark}},\ and\
  \bibinfo {author} {\bibfnamefont {W.}~\bibnamefont {Wurth}},\ }\bibfield
  {title} {\bibinfo {title} {Short-wavelength free-electron laser sources and
  science: a review},\ }\href {https://doi.org/10.1088/1361-6633/aa7cca}
  {\bibfield  {journal} {\bibinfo  {journal} {Reports on Progress in Physics}\
  }\textbf {\bibinfo {volume} {80}},\ \bibinfo {pages} {115901} (\bibinfo
  {year} {2017})}\BibitemShut {NoStop}%
\bibitem [{\citenamefont {Huang}\ \emph {et~al.}(2021)\citenamefont {Huang},
  \citenamefont {Deng}, \citenamefont {Liu}, \citenamefont {Wang},\ and\
  \citenamefont {Zhao}}]{Huang2021}%
  \BibitemOpen
  \bibfield  {author} {\bibinfo {author} {\bibfnamefont {N.}~\bibnamefont
  {Huang}}, \bibinfo {author} {\bibfnamefont {H.}~\bibnamefont {Deng}},
  \bibinfo {author} {\bibfnamefont {B.}~\bibnamefont {Liu}}, \bibinfo {author}
  {\bibfnamefont {D.}~\bibnamefont {Wang}},\ and\ \bibinfo {author}
  {\bibfnamefont {Z.}~\bibnamefont {Zhao}},\ }\bibfield  {title} {\bibinfo
  {title} {{Features and futures of X-ray free-electron lasers}},\ }\href
  {https://doi.org/10.1016/j.xinn.2021.100097} {\bibfield  {journal} {\bibinfo
  {journal} {The Innovation}\ }\textbf {\bibinfo {volume} {2}},\ \bibinfo
  {pages} {100097} (\bibinfo {year} {2021})}\BibitemShut {NoStop}%
\bibitem [{\citenamefont {Johnson}\ \emph {et~al.}(2019)\citenamefont
  {Johnson}, \citenamefont {Avni}, \citenamefont {Larsen}, \citenamefont
  {Austin},\ and\ \citenamefont {Marangos}}]{Johnson2019}%
  \BibitemOpen
  \bibfield  {author} {\bibinfo {author} {\bibfnamefont {A.~S.}\ \bibnamefont
  {Johnson}}, \bibinfo {author} {\bibfnamefont {T.}~\bibnamefont {Avni}},
  \bibinfo {author} {\bibfnamefont {E.~W.}\ \bibnamefont {Larsen}}, \bibinfo
  {author} {\bibfnamefont {D.~R.}\ \bibnamefont {Austin}},\ and\ \bibinfo
  {author} {\bibfnamefont {J.~P.}\ \bibnamefont {Marangos}},\ }\bibfield
  {title} {\bibinfo {title} {{Attosecond soft X-ray high harmonic
  generation}},\ }\href {https://doi.org/10.1098/rsta.2017.0468} {\bibfield
  {journal} {\bibinfo  {journal} {Philosophical Transactions of the Royal
  Society A: Mathematical, Physical and Engineering Sciences}\ }\textbf
  {\bibinfo {volume} {377}},\ \bibinfo {pages} {20170468} (\bibinfo {year}
  {2019})}\BibitemShut {NoStop}%
\bibitem [{\citenamefont {Huang}\ \emph {et~al.}(2016)\citenamefont {Huang}
  \emph {et~al.}}]{Huang2016}%
  \BibitemOpen
  \bibfield  {author} {\bibinfo {author} {\bibfnamefont {S.}~\bibnamefont
  {Huang}} \emph {et~al.},\ }\bibfield  {title} {\bibinfo {title} {Generation
  of sub-terawatt-attosecond pulses in a soft-x-ray self-seeded free-electron
  laser},\ }\href {https://doi.org/10.1103/PhysRevAccelBeams.19.080702}
  {\bibfield  {journal} {\bibinfo  {journal} {Physical Review Accelerators and
  Beams}\ }\textbf {\bibinfo {volume} {19}},\ \bibinfo {pages} {080702}
  (\bibinfo {year} {2016})}\BibitemShut {NoStop}%
\bibitem [{\citenamefont {Georges}\ \emph {et~al.}(1996)\citenamefont
  {Georges}, \citenamefont {Kotliar}, \citenamefont {Krauth},\ and\
  \citenamefont {Rozenberg}}]{Georges1996}%
  \BibitemOpen
  \bibfield  {author} {\bibinfo {author} {\bibfnamefont {A.}~\bibnamefont
  {Georges}}, \bibinfo {author} {\bibfnamefont {G.}~\bibnamefont {Kotliar}},
  \bibinfo {author} {\bibfnamefont {W.}~\bibnamefont {Krauth}},\ and\ \bibinfo
  {author} {\bibfnamefont {M.~J.}\ \bibnamefont {Rozenberg}},\ }\bibfield
  {title} {\bibinfo {title} {Dynamical mean-field theory of strongly correlated
  fermion systems and the limit of infinite dimensions},\ }\href
  {https://doi.org/10.1103/RevModPhys.68.13} {\bibfield  {journal} {\bibinfo
  {journal} {Rev. Mod. Phys.}\ }\textbf {\bibinfo {volume} {68}},\ \bibinfo
  {pages} {13} (\bibinfo {year} {1996})}\BibitemShut {NoStop}%
\bibitem [{\citenamefont {Eckstein}\ \emph {et~al.}(2009)\citenamefont
  {Eckstein}, \citenamefont {Kollar},\ and\ \citenamefont
  {Werner}}]{Eckstein2009}%
  \BibitemOpen
  \bibfield  {author} {\bibinfo {author} {\bibfnamefont {M.}~\bibnamefont
  {Eckstein}}, \bibinfo {author} {\bibfnamefont {M.}~\bibnamefont {Kollar}},\
  and\ \bibinfo {author} {\bibfnamefont {P.}~\bibnamefont {Werner}},\
  }\bibfield  {title} {\bibinfo {title} {Thermalization after an interaction
  quench in the hubbard model},\ }\href
  {https://doi.org/10.1103/PhysRevLett.103.056403} {\bibfield  {journal}
  {\bibinfo  {journal} {Phys. Rev. Lett.}\ }\textbf {\bibinfo {volume} {103}},\
  \bibinfo {pages} {056403} (\bibinfo {year} {2009})}\BibitemShut {NoStop}%
\bibitem [{\citenamefont {Amaricci}\ \emph {et~al.}(2015)\citenamefont
  {Amaricci}, \citenamefont {Weber}, \citenamefont {Capone}, \citenamefont
  {Kotliar},\ and\ \citenamefont {Millis}}]{Amaricci2015}%
  \BibitemOpen
  \bibfield  {author} {\bibinfo {author} {\bibfnamefont {A.}~\bibnamefont
  {Amaricci}}, \bibinfo {author} {\bibfnamefont {C.}~\bibnamefont {Weber}},
  \bibinfo {author} {\bibfnamefont {M.}~\bibnamefont {Capone}}, \bibinfo
  {author} {\bibfnamefont {G.}~\bibnamefont {Kotliar}},\ and\ \bibinfo {author}
  {\bibfnamefont {A.~J.}\ \bibnamefont {Millis}},\ }\bibfield  {title}
  {\bibinfo {title} {Exploring the energy landscape of the non-equilibrium
  hubbard model},\ }\href {https://doi.org/10.1103/PhysRevLett.114.246402}
  {\bibfield  {journal} {\bibinfo  {journal} {Phys. Rev. Lett.}\ }\textbf
  {\bibinfo {volume} {114}},\ \bibinfo {pages} {246402} (\bibinfo {year}
  {2015})}\BibitemShut {NoStop}%
\bibitem [{\citenamefont {Werner}\ \emph {et~al.}(2012)\citenamefont {Werner},
  \citenamefont {Tsuji},\ and\ \citenamefont {Eckstein}}]{Werner2012}%
  \BibitemOpen
  \bibfield  {author} {\bibinfo {author} {\bibfnamefont {P.}~\bibnamefont
  {Werner}}, \bibinfo {author} {\bibfnamefont {N.}~\bibnamefont {Tsuji}},\ and\
  \bibinfo {author} {\bibfnamefont {M.}~\bibnamefont {Eckstein}},\ }\bibfield
  {title} {\bibinfo {title} {Nonthermal symmetry-broken states in the strongly
  interacting hubbard model},\ }\href
  {https://doi.org/10.1103/PhysRevB.86.205101} {\bibfield  {journal} {\bibinfo
  {journal} {Phys. Rev. B}\ }\textbf {\bibinfo {volume} {86}},\ \bibinfo
  {pages} {205101} (\bibinfo {year} {2012})}\BibitemShut {NoStop}%
\bibitem [{\citenamefont {Schollwöck}(2011)}]{Schollwock2011}%
  \BibitemOpen
  \bibfield  {author} {\bibinfo {author} {\bibfnamefont {U.}~\bibnamefont
  {Schollwöck}},\ }\bibfield  {title} {\bibinfo {title} {The density-matrix
  renormalization group in the age of matrix product states},\ }\href
  {https://doi.org/10.1016/j.aop.2010.09.012} {\bibfield  {journal} {\bibinfo
  {journal} {Annals of Physics}\ }\textbf {\bibinfo {volume} {326}},\ \bibinfo
  {pages} {96} (\bibinfo {year} {2011})}\BibitemShut {NoStop}%
\bibitem [{\citenamefont {Kollath}\ \emph {et~al.}(2007)\citenamefont
  {Kollath}, \citenamefont {Läuchli},\ and\ \citenamefont
  {Altman}}]{Kollath2007}%
  \BibitemOpen
  \bibfield  {author} {\bibinfo {author} {\bibfnamefont {C.}~\bibnamefont
  {Kollath}}, \bibinfo {author} {\bibfnamefont {A.~M.}\ \bibnamefont
  {Läuchli}},\ and\ \bibinfo {author} {\bibfnamefont {E.}~\bibnamefont
  {Altman}},\ }\bibfield  {title} {\bibinfo {title} {Quench dynamics and
  non-equilibrium phase diagram of the bose-hubbard model},\ }\href
  {https://doi.org/10.1103/PhysRevLett.98.180601} {\bibfield  {journal}
  {\bibinfo  {journal} {Phys. Rev. Lett.}\ }\textbf {\bibinfo {volume} {98}},\
  \bibinfo {pages} {180601} (\bibinfo {year} {2007})}\BibitemShut {NoStop}%
\bibitem [{\citenamefont {Barmettler}\ \emph {et~al.}(2008)\citenamefont
  {Barmettler}, \citenamefont {Punk}, \citenamefont {Gritsev}, \citenamefont
  {Demler},\ and\ \citenamefont {Altman}}]{Barmettler2008}%
  \BibitemOpen
  \bibfield  {author} {\bibinfo {author} {\bibfnamefont {P.}~\bibnamefont
  {Barmettler}}, \bibinfo {author} {\bibfnamefont {M.}~\bibnamefont {Punk}},
  \bibinfo {author} {\bibfnamefont {V.}~\bibnamefont {Gritsev}}, \bibinfo
  {author} {\bibfnamefont {E.}~\bibnamefont {Demler}},\ and\ \bibinfo {author}
  {\bibfnamefont {E.}~\bibnamefont {Altman}},\ }\bibfield  {title} {\bibinfo
  {title} {Propagation front of correlations in an interacting bose gas},\
  }\href {https://doi.org/10.1103/PhysRevA.78.012330} {\bibfield  {journal}
  {\bibinfo  {journal} {Phys. Rev. A}\ }\textbf {\bibinfo {volume} {78}},\
  \bibinfo {pages} {012330} (\bibinfo {year} {2008})}\BibitemShut {NoStop}%
\bibitem [{\citenamefont {Daley}\ \emph {et~al.}(2004)\citenamefont {Daley},
  \citenamefont {Kollath}, \citenamefont {Schollwöck},\ and\ \citenamefont
  {Vidal}}]{Daley2004}%
  \BibitemOpen
  \bibfield  {author} {\bibinfo {author} {\bibfnamefont {A.~J.}\ \bibnamefont
  {Daley}}, \bibinfo {author} {\bibfnamefont {C.}~\bibnamefont {Kollath}},
  \bibinfo {author} {\bibfnamefont {U.}~\bibnamefont {Schollwöck}},\ and\
  \bibinfo {author} {\bibfnamefont {G.}~\bibnamefont {Vidal}},\ }\bibfield
  {title} {\bibinfo {title} {Time-dependent density-matrix
  renormalization-group using adaptive effective hilbert spaces},\ }\href
  {https://doi.org/10.1088/1742-5468/2004/04/P04005} {\bibfield  {journal}
  {\bibinfo  {journal} {Journal of Statistical Mechanics: Theory and
  Experiment}\ }\textbf {\bibinfo {volume} {2004}},\ \bibinfo {pages} {P04005}
  (\bibinfo {year} {2004})}\BibitemShut {NoStop}%
\end{thebibliography}%
\end{document}

%\title{Supplemental Material for \\ Unveiling Universal Photoresponse Dynamics in a Cuprate Superconductor Excited by Near-Infrared and EUV Light}
\title{Supplemental Material for \\ Energy density driven ultrafast electronic excitations in a cuprate
   superconductor}

\author{Alessandra Milloch}
\affiliation{Department of Mathematics and Physics, Università Cattolica del Sacro Cuore, Brescia I-25133, Italy}
\affiliation{ILAMP (Interdisciplinary Laboratories for Advanced
Materials Physics), Università Cattolica del Sacro Cuore, Brescia I-25133, Italy}

\author{Francesco Proietto}
\affiliation{Department of Mathematics and Physics, Università Cattolica del Sacro Cuore, Brescia I-25133, Italy}
\affiliation{ILAMP (Interdisciplinary Laboratories for Advanced
Materials Physics), Università Cattolica del Sacro Cuore, Brescia I-25133, Italy}
\affiliation{Department of Physics and Astronomy, KU Leuven, B-3001 Leuven, Belgium}

\author{Naman Agarwal}
\affiliation{Elettra - Sincrotrone Trieste S.C.p.A., Trieste, Italy}

\author{Laura Foglia}
\affiliation{Elettra - Sincrotrone Trieste S.C.p.A., Trieste, Italy}

\author{Riccardo Mincigrucci} 
\affiliation{Elettra - Sincrotrone Trieste S.C.p.A., Trieste, Italy}

\author{Genda Gu}
\affiliation{Condensed Matter Physics and Materials Science Division, Brookhaven National Laboratory,
Upton, New York 11973, USA}

\author{Claudio Giannetti}
\affiliation{Department of Mathematics and Physics, Università Cattolica del Sacro Cuore, Brescia I-25133, Italy}
\affiliation{ILAMP (Interdisciplinary Laboratories for Advanced
Materials Physics), Università Cattolica del Sacro Cuore, Brescia I-25133, Italy}
\affiliation{CNR-INO (National Institute of Optics), via Branze 45, 25123 Brescia, Italy}

\author{Federico Cilento}
\affiliation{Elettra - Sincrotrone Trieste S.C.p.A., Trieste, Italy}

\author{Filippo Bencivenga}
\affiliation{Elettra - Sincrotrone Trieste S.C.p.A., Trieste, Italy}

\author{Fulvio Parmigiani}
\email[]{fulvio.parmigiani@elettra.eu}
\affiliation{Dipartimento di Fisica, Università degli Studi di Trieste, Trieste, Italy}
\affiliation{Elettra - Sincrotrone Trieste S.C.p.A., Trieste, Italy}

\maketitle

\section{Experimental setup}
\begin{figure*}[h!]
\centering
\includegraphics[width=12cm]{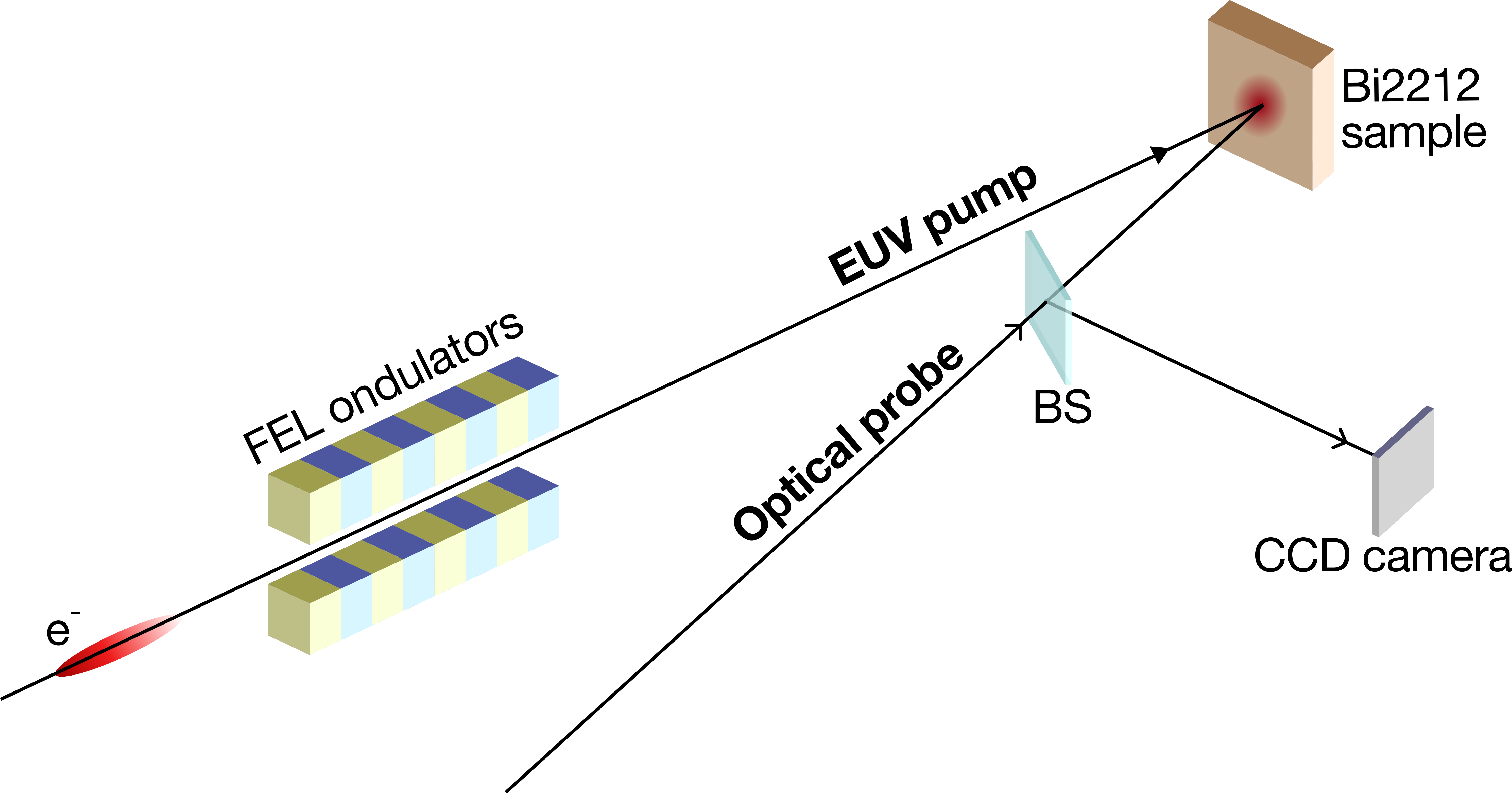}
\caption{Sketch of the experimental setup for FEL-pump, optical probe measurements. }
\label{fig: setup}
\end{figure*}

%\newpage
\section{Reflectivity model}
\begin{figure*}[h!]
\centering
\includegraphics[width=13cm]{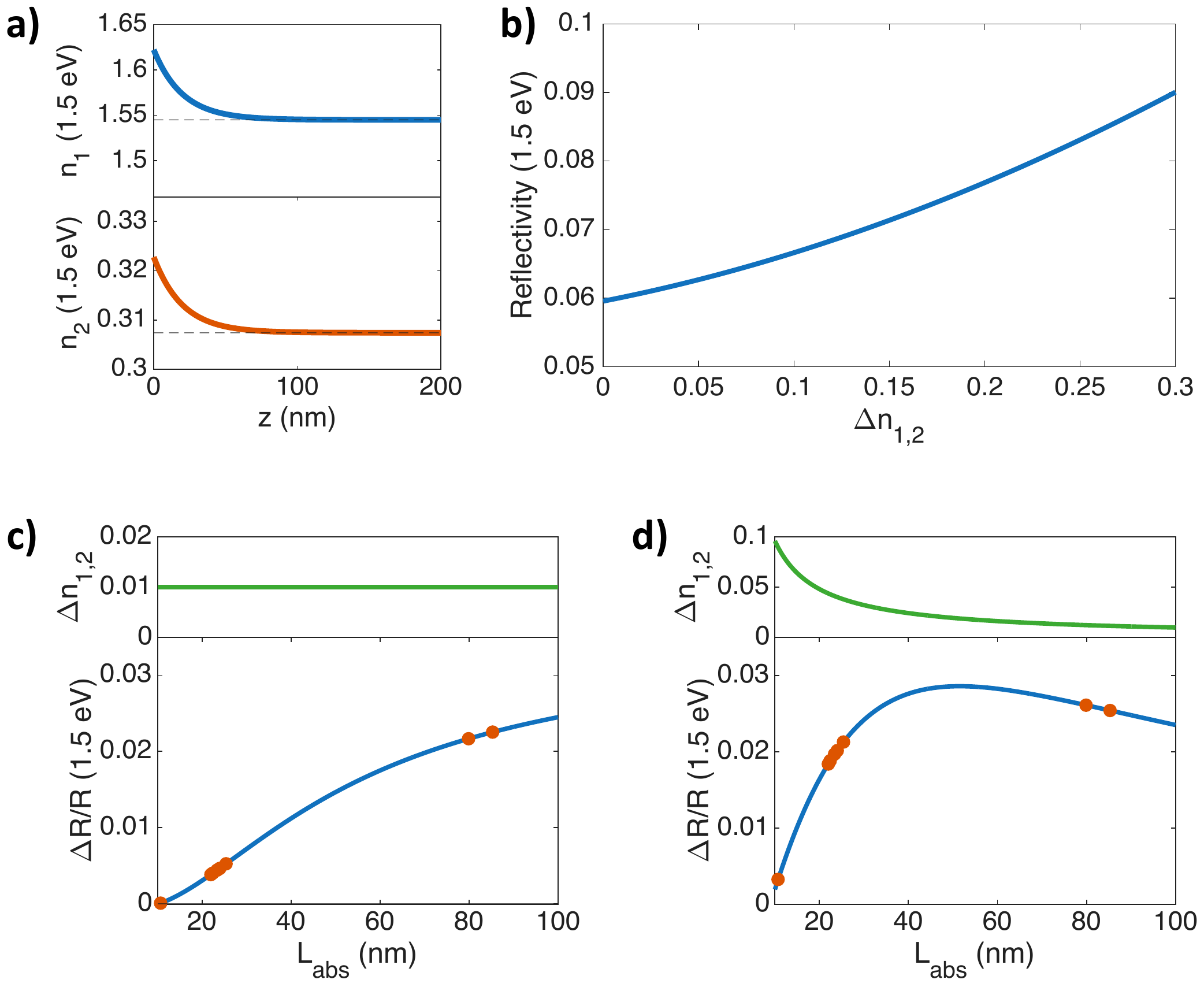}
\caption{a) Exponential profiles of the refractive index in the sample after pump excitation modelled through Eq. \ref{eq: ref index exp} and plotted for $\Delta n = 0.05$ and $L_\mathrm{abs} = 20$~nm. Dashed gray lines represent the equilibrium values of $n_1$ and $n_2$ at 1.5~eV (taken from Ref. \citenum{giannetti2011revealing}). b) 1.5 eV reflectivity for a 3~\textmu m Bi2212 sample with refractive index decaying exponentially (considering $L_\mathrm{abs} = 50$~bn), plotted as a function of the amplitude of the refractive index increase $\Delta n$. c) Change in reflectivity at 1.5 eV compared to equilibrium (homogeneous medium), for a refractive index $\Delta n = 0.01$ (top panel) at the surface, decaying exponentially over $L_\mathrm{abs}$. d) Same as c), but considering a different $\Delta n$ at the surface for different values of $L_\mathrm{abs}$, in order to compare the effect for the same average increase of refractive index over the volume. Red markers indicate the values of $L_\mathrm{abs}$ corresponding to the FEL pump photon energies employed in the experiment. }
\label{fig: trasfer matrix}
\end{figure*}

In order to account for the exponential spatial profile of the pump excitation, we model the excited sample as a medium where the refractive index $n$ varies exponentially with sample depth $z$, that is: 
\begin{equation}
    n = n_0 (1+\Delta n e^{-z/L_\mathrm{abs}})
    \label{eq: ref index exp}
\end{equation}
where $n_0$ is the equilibrium refractive index, $\Delta n$ represents the pump-induced change in the refractive index at the surface, and $L_\mathrm{abs}$ is the pump absorption length. %Considering an overall sample thickness of 3~\textmu m (bulk limit) and describing it as a layered material composed of 1~nm layers where the refractive index varies layer by layer as given by Eq. \ref{eq: ref index exp}, the reflectivity of the excited sample can be computed using the transfer matrix method. 
Assuming a total sample thickness of 3~\textmu m (bulk limit), the medium is modeled as a stack of 1~nm layers, with the refractive index of each layer depends on the layer position $z$ according to Eq.~\ref{eq: ref index exp}. The overall reflectivity is then computed using the transfer-matrix method. 
Considering an increase in both the real ($n_1$) and imaginary ($n_2$) parts of the refractive index ($\Delta n > 0$) as sketched in Fig. \ref{fig: trasfer matrix}a, the reflectivity at 1.5 eV is estimated to increase as compared to the equilibrium reflectivity ($\Delta n$ = 0) and scales with $\Delta n$ as shown in Fig. \ref{fig: trasfer matrix}b. \\
We now aim to study how the pump-induced reflectivity increase depends on the pump absorption length. Fig. \ref{fig: trasfer matrix}c shows the estimated $\Delta R/R$ as a function of $L_\mathrm{abs}$ considering $\Delta n = 0.01$. Since in the experiment we compare data acquired at similar absorbed energy density ($I_{\mathrm{exc}}$) rather than incident fluence ($F$), we now consider a scenario in which $\Delta n$ scales inversely with $L_\mathrm{abs}$ (see Fig. \ref{fig: trasfer matrix}d, top panel), ensuring the same \textit{average} refractive index change for all  $L_\mathrm{abs}$. The resulting dependence of 
$\Delta R/R$ is shown in Fig.~\ref{fig: trasfer matrix}d. 
The red markers indicate the values of $L_\mathrm{abs}$ corresponding to our experimental data and provide the normalization factor that can be employed to compare the experimental pump-probe signal across different FEL pump photon energies. Specifically, we define the normalization factor as 
\begin{equation}
    r = \frac{\Delta R (L_\mathrm{abs}^\mathrm{FEL})}{\Delta R (L_\mathrm{abs}^\mathrm
    {opt})},
\end{equation}
that is, as the ratio between the transient reflectivity signal amplitude expected for FEL pump with short absorption length (red markers in Fig. \ref{fig: trasfer matrix}d) and that for the optical case of near-IR pump ($\Delta R/R (L_\mathrm{abs}^\mathrm{opt} = 166~\text{nm})$ = 0.0168). Employment of the reflectivity model to compare experimental signal amplitudes at 1.5~eV probe showed that accounting for the normalization factor $r = \Delta R(L_{\mathrm{abs}}^{\mathrm{FEL}}) / \Delta R(L_{\mathrm{abs}}^{\mathrm{opt}})$  results in comparable signal amplitudes at all pump photon energies.
%\color{blue}{The reflectivity model was validated by comparison to experimental amplitude ratios at 1.5~eV, ensuring that the normalization factor $r = \Delta R(L_{\mathrm{abs}}^{\mathrm{FEL}}) / \Delta R(L_{\mathrm{abs}}^{\mathrm{opt}})$ reproduces the observed trends within 10\%.}

%\newpage 
\section{Supplementary data}

\begin{figure*}[]
\centering
\includegraphics[width=13cm]{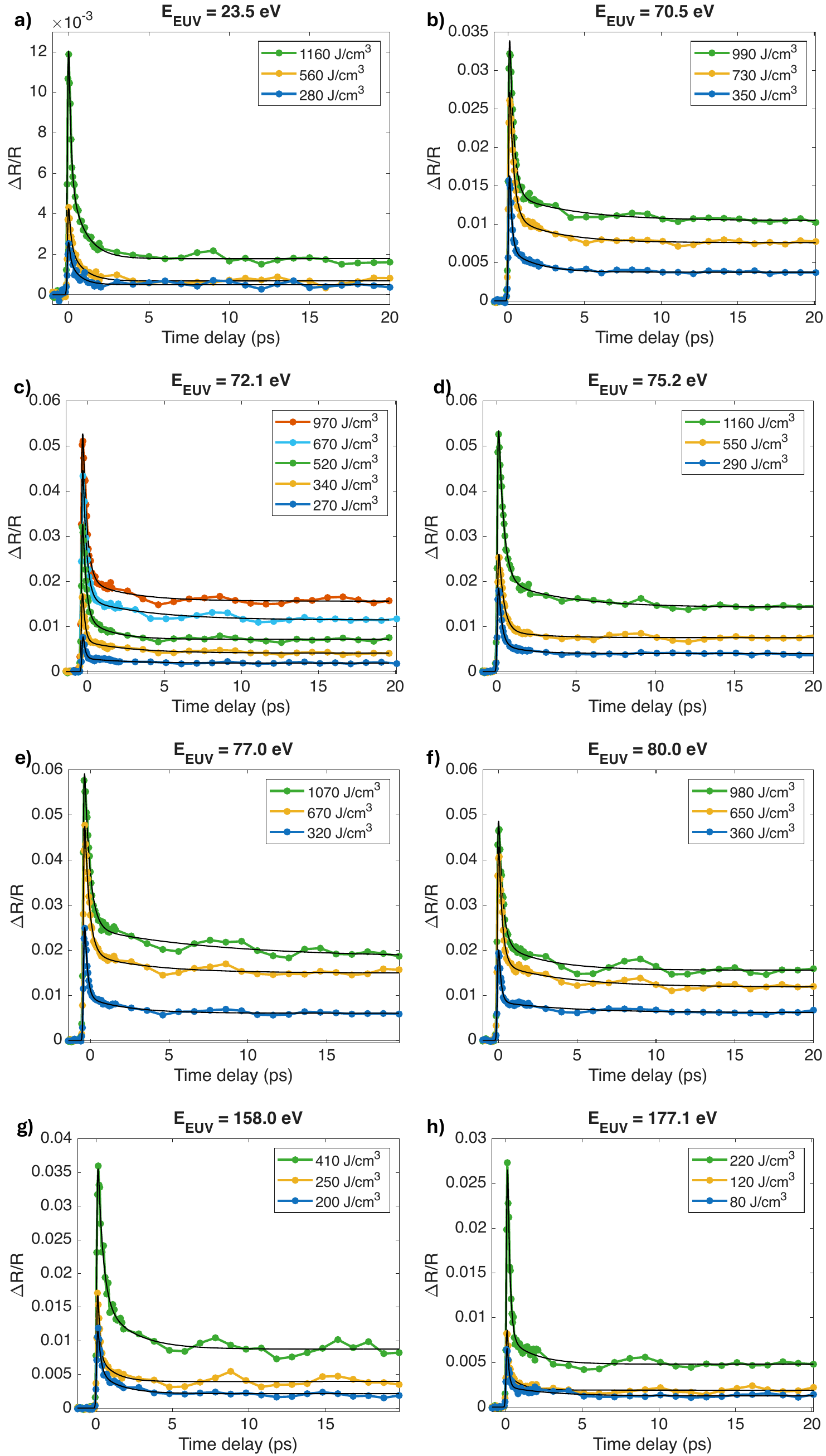}
\caption{Fluence dependence of the transient reflectivity signal at room temperature, measured with FEL excitation at different photon energies $E_{\mathrm{EUV}}$. Black solid lines represent multi-exponential fits to the measured $\Delta R/R$ dynamics.}
\label{fig: RT dynamics}
\end{figure*}

\begin{figure*}[]
\centering
\includegraphics[width=10cm]{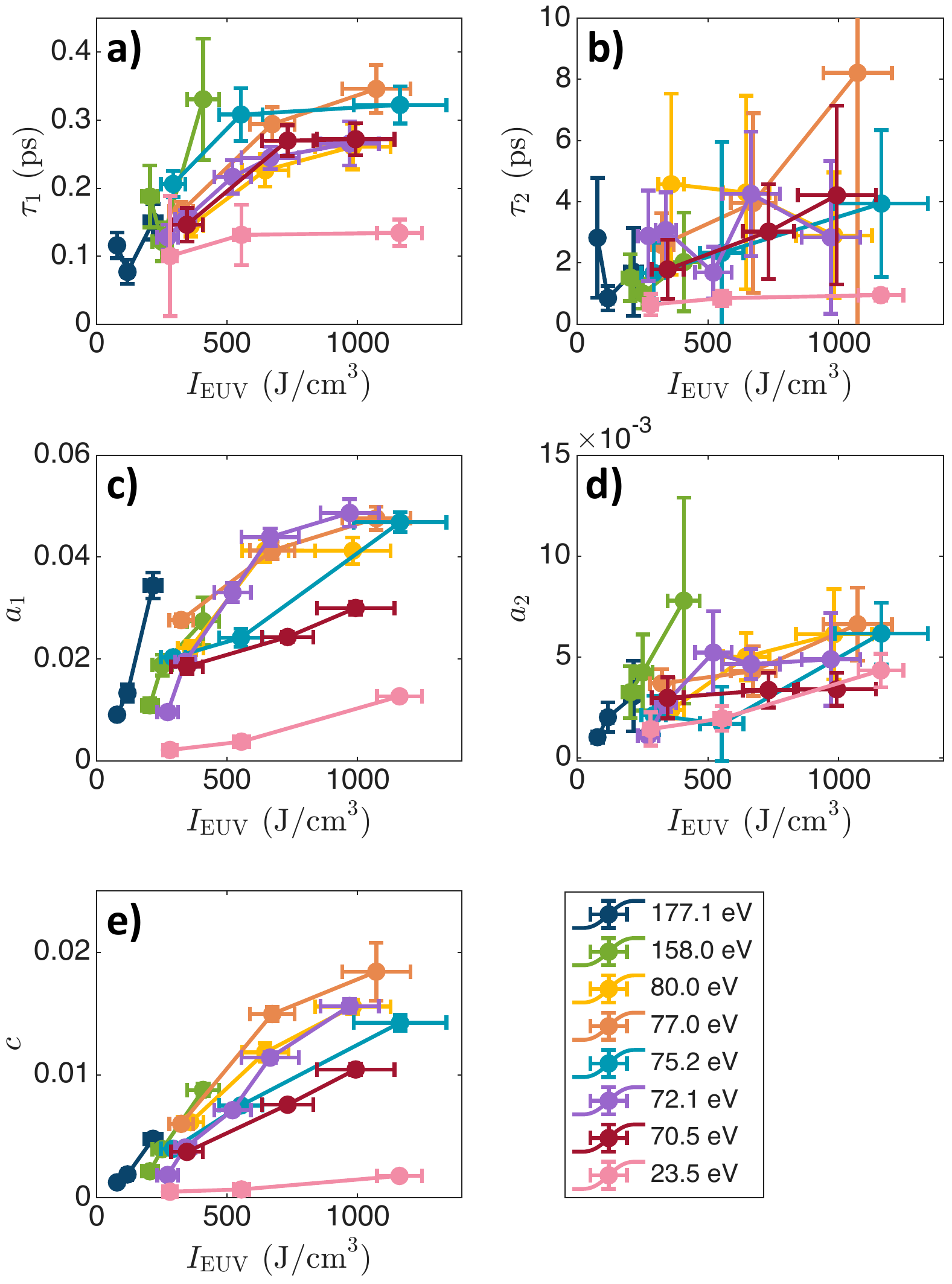}
\caption{Fit parameters extracted from room-temperature FEL-pump, optical-probe measurements. Parameters are obtained from multi-exponential fits to the $\Delta R/R$ dynamics.
a,b) decay times, c,d,e) amplitude parameters. 
All quantities are plotted as a function of excitation energy density $I_{\mathrm{EUV}}$ for the different pump photon energies employed.}
\label{fig: param RT not norm}
\end{figure*}

\begin{figure*}[]
\centering
\includegraphics[width=13cm]{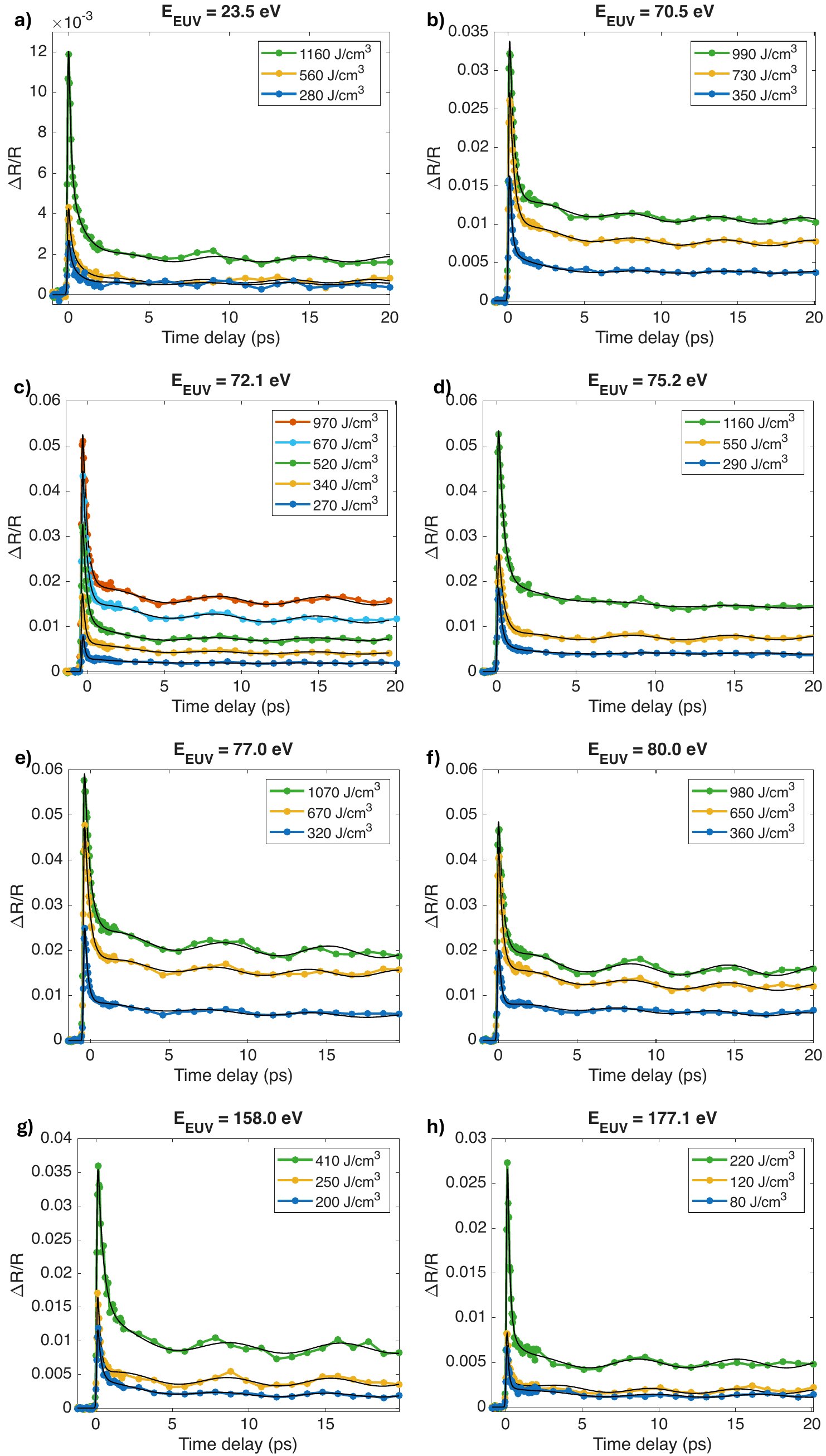}
\caption{Room-temperature $\Delta R/R$ dynamics including oscillatory fit. Same data as in Fig. S2, now fitted with a function consisting of a double-exponential decay plus an additional oscillatory term to account for the $\sim6$~s oscillation observed in the 2–20~ps time window. Black solid lines indicate the fit.}
\label{fig: fit osc}
\end{figure*}

\begin{figure*}[]
\centering
\includegraphics[width=14cm]{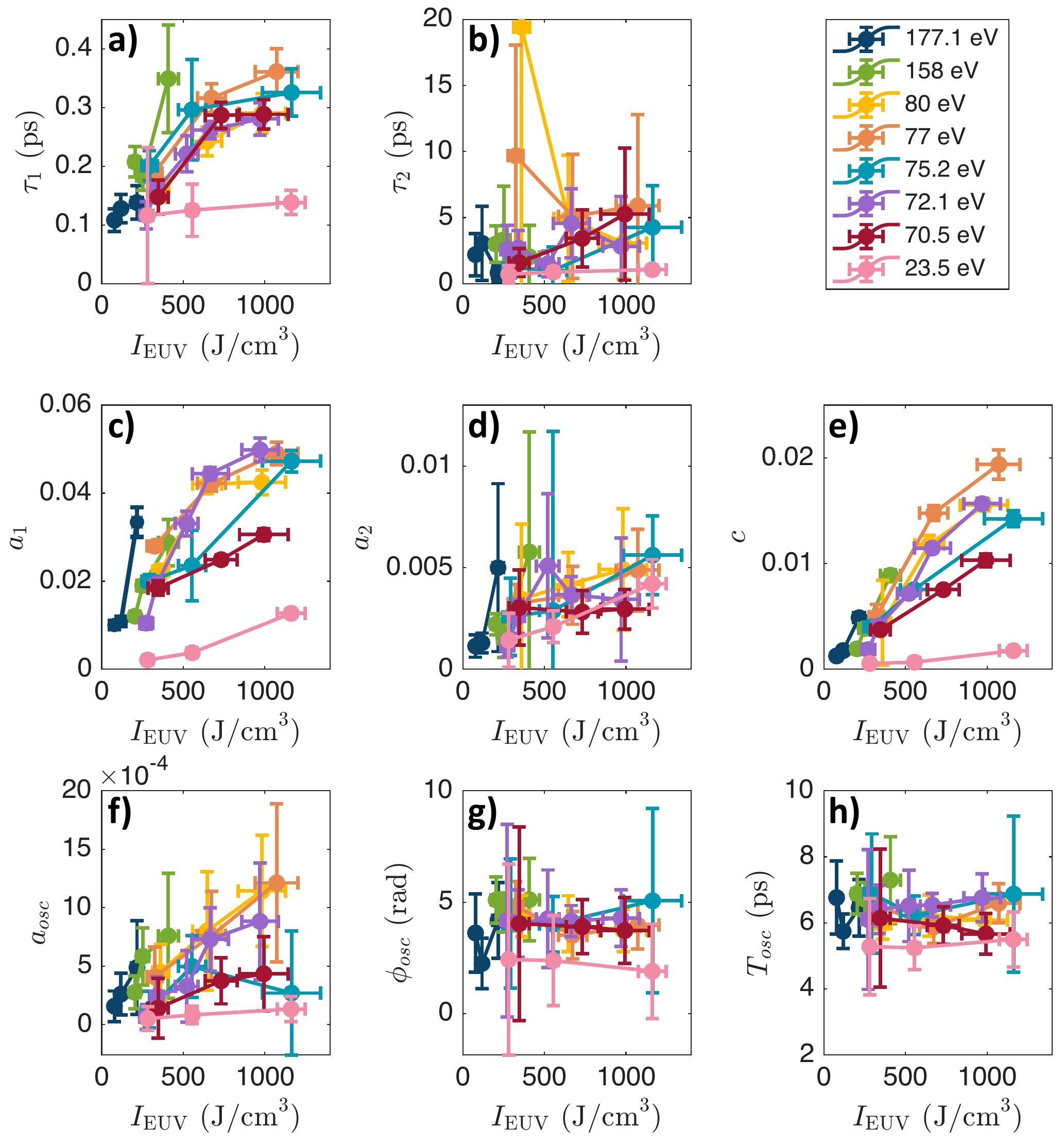}
\caption{Fit parameters including oscillatory component. Output parameters from the fits shown in Fig. S4, where an oscillatory cosine term is added to a double-exponential decay function. a,b) Decay times. c,d,e) Exponential amplitude parameters. f) Amplitude of the oscillatory contribution. g) Phase of the cosine term. h) Oscillation period. All parameters are plotted as a function of excitation energy density $I_{\mathrm{EUV}}$ for each pump photon energy employed.}
\label{fig: fit param osc}
\end{figure*}

\begin{figure*}[]
\centering
\includegraphics[width=9cm]{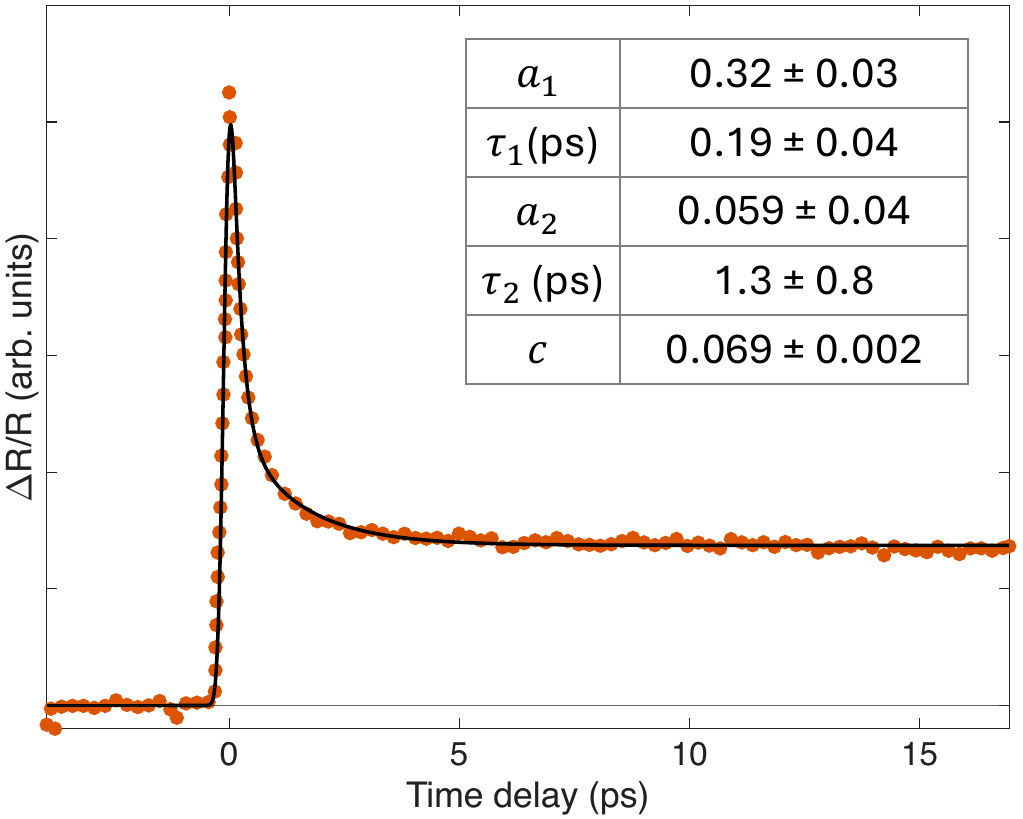}
\caption{Room-temperature optical-pump transient reflectivity. Transient reflectivity trace measured at room temperature with a 1.5~eV optical pump, adapted from Ref.~\citenum{giannetti2009discontinuity}. The black line shows the double-exponential fit, with the corresponding fit parameters summarized in the inset table. }
\label{fig: optics RT}
\end{figure*}

\begin{figure*}[]
\centering
\includegraphics[width=15cm]{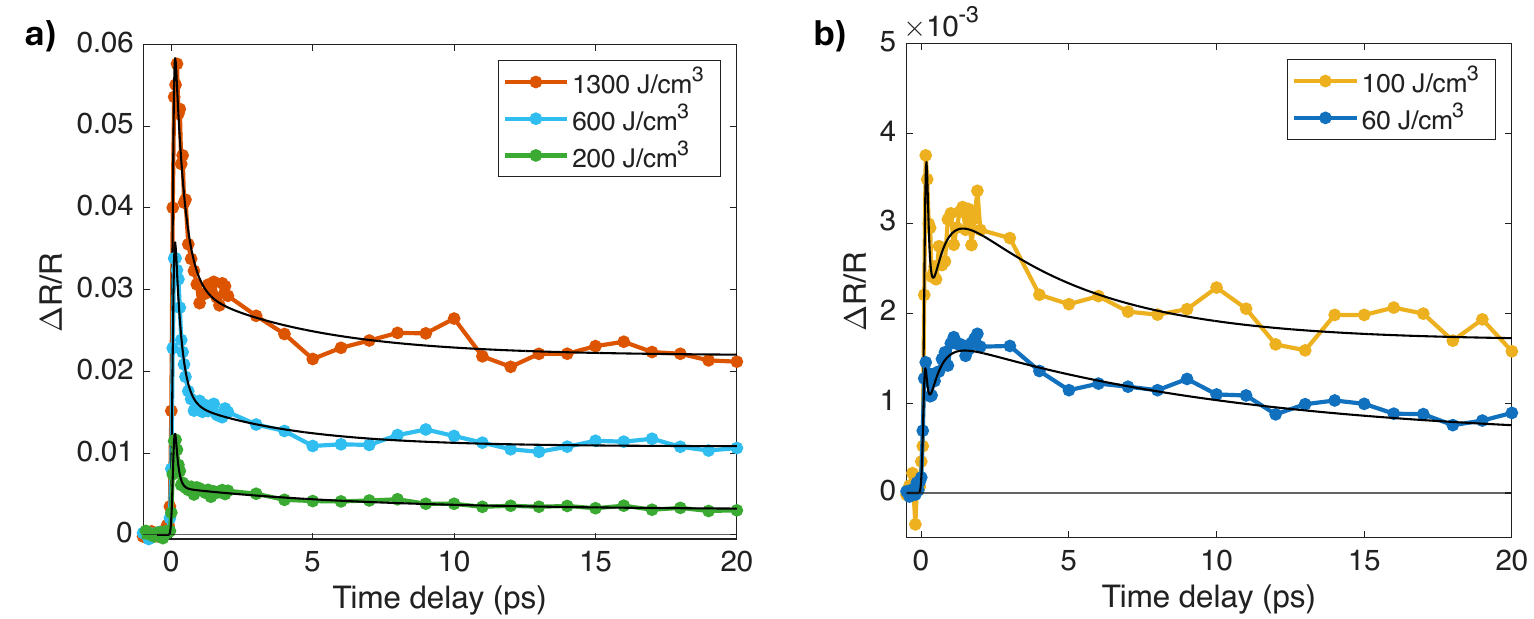}
\caption{Fluence dependence of the transient reflectivity signal at $T = 30$~K, measured with FEL excitation at $E_{\mathrm{EUV}} = 77$~eV. Black solid lines show multi-exponential fits to the measured $\Delta R/R$ dynamics. a) At high excitation fluence, the response is well described by a two-exponential decay. b) At low excitation density, an additional exponential component is required to reproduce the delayed build-up characteristic of a boson bottleneck effect.}
\label{fig: fit low T}
\end{figure*}

\begin{figure*}[]
\centering
\includegraphics[width=18cm]{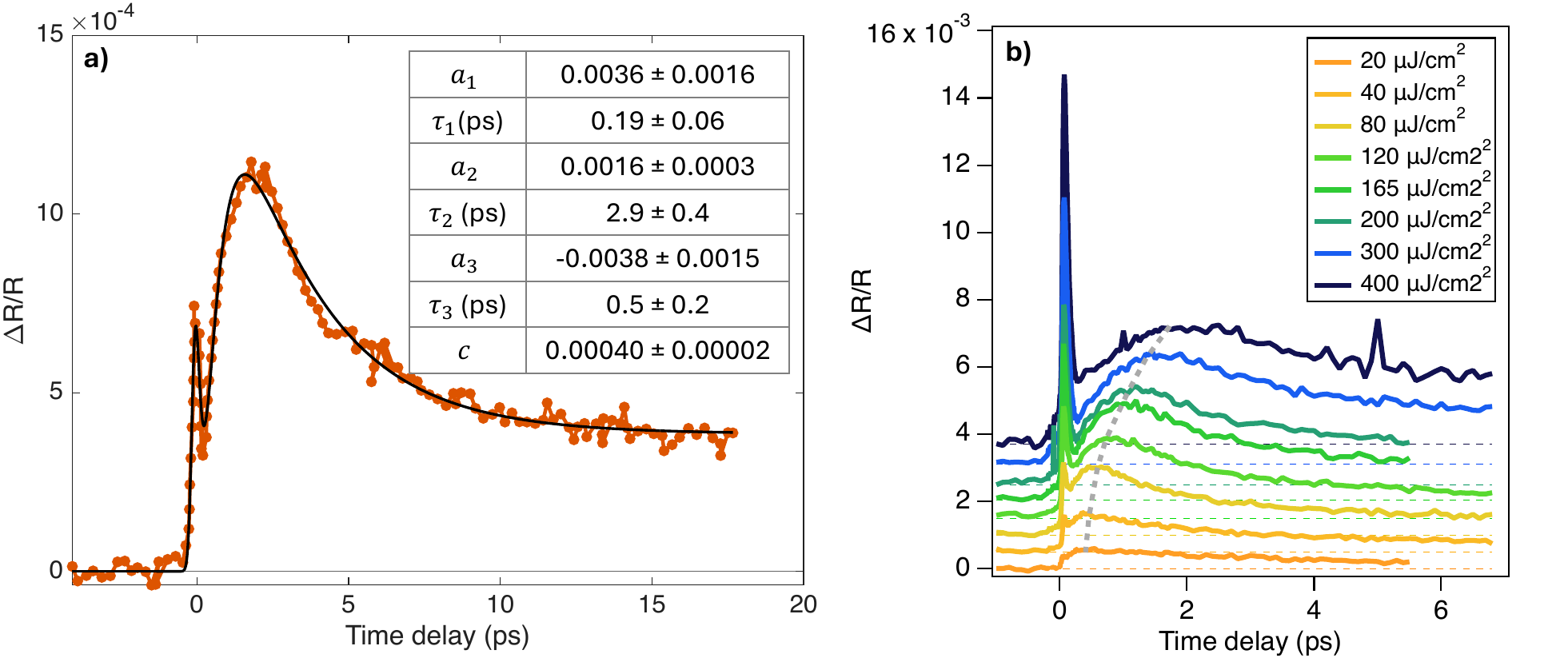}
\caption{Optical-pump transient reflectivity below $T_c$.
a) Transient reflectivity trace adapted from Ref.\citenum{giannetti2009discontinuity}, measured with a 1.5~eV optical pump and probe at a fluence of $285$~$\textmu$J/cm$^2$ and temperature $T = 20$~K. The black line shows the three-exponential fit, with the extracted parameters summarized in the inset table.
b) Fluence-dependent transient reflectivity dynamics measured at $T = 60$~K on a Y-substituted optimally-doped Bi2212 sample (\ch{Bi_2Sr_2Ca_{0.92}Y_{0.08}Cu_2O_{8+$\delta$}}, $\delta \simeq  0.16$, $T_c = 95$~K). The experiment employed a broadband pump pulse (1.3–1.8~eV, $\sim$30~fs time duration) and a 1.65~eV probe. The gray dashed line is a guide to the eye highlighting the delayed peak associated with partial suppression of the superconducting condensate, which shifts to longer times with increasing fluence, consistent with Ref.~\citenum{giannetti2009discontinuity}.}
\label{fig: optics LT}
\end{figure*}

\clearpage
\bibliography{Refs}